\documentclass{aa}  

\usepackage{graphicx}
\usepackage[varg]{txfonts}
\usepackage{lipsum}
\usepackage{natbib,twoopt}
\usepackage[breaklinks=true]{hyperref} %% to avoid \citeads line fills
\bibpunct{(}{)}{;}{a}{}{,} %% natbib format for A&A and ApJ
\makeatletter
\newcommandtwoopt{\citeads}[3][][]{\href{http://adsabs.harvard.edu/abs/#3}%
{\def\hyper@linkstart##1##2{}%
\let\hyper@linkend\@empty\citealp[#1][#2]{#3}}}
\newcommandtwoopt{\citepads}[3][][]{\href{http://adsabs.harvard.edu/abs/#3}%
{\def\hyper@linkstart##1##2{}%
\let\hyper@linkend\@empty\citep[#1][#2]{#3}}}
\newcommandtwoopt{\citetads}[3][][]{\href{http://adsabs.harvard.edu/abs/#3}%
{\def\hyper@linkstart##1##2{}%
\let\hyper@linkend\@empty\citet[#1][#2]{#3}}}
\newcommandtwoopt{\citeyearads}[3][][]%
{\href{http://adsabs.harvard.edu/abs/#3}
{\def\hyper@linkstart##1##2{}%
\let\hyper@linkend\@empty\citeyear[#1][#2]{#3}}}
\makeatother

\begin{document}

   \title{An internal heating mechanism operating in ultra-short-period planets orbiting magnetically active stars}

%   \subtitle{I. Overviewing the $\kappa$-mechanism}
\titlerunning{An internal heating mechanism in ultra-short-period planets }

   \author{A.~F.~Lanza
%          \inst{1}
          }
   \institute{INAF-Osservatorio Astrofisico di Catania, Via S.~Sofia, 78 - 95123 Catania, Italy    \\          
   \email{antonino.lanza@inaf.it}}
         
   \date{Received ... ; accepted ... }

% \abstract{}{}{}{}{} 
% 5 {} token are mandatory
 
  \abstract
   {Rocky planets with orbital periods shorter than $\sim 1$~day have been discovered by the method of transits and their study can provide information on Earth-like planets not available from bodies on longer period orbits.}
   {A new mechanism for the internal heating of such ultra-short-period planets is proposed based on the gravitational perturbation produced by a non-axisymmetric quadrupole moment of their host stars. Such a quadrupole is due to the magnetic flux tubes in the stellar convection zone, unevenly distributed in longitude and persisting for many stellar rotations as observed in young late-type stars. }
   {The rotation period of the host star evolves from its shortest value on the zero-age main sequence (ZAMS) to longer periods due to the loss of angular momentum through a magnetized wind. If the stellar rotation period comes close to twice the orbital period of the planet, the quadrupole leads to a spin-orbit resonance that excites oscillations of the star-planet separation. As a consequence, a strong tidal dissipation is produced inside the planet that converts the energy of the oscillations into internal heat. The total heat released inside the planet  scales as $a^{-8}$, where $a$ is the orbit semimajor axis, and is largely independent of the details of the planetary internal dissipation or the lifetime of the stellar magnetic flux tubes.}
   {We illustrate the operation of the mechanism by modelling the evolution of the stellar rotation and of the innermost planetary orbit under the action of the stellar wind and the tides in the cases of CoRoT-7, Kepler-78, and K2-141 whose present orbital periods range between 0.28 and 0.85 days. If the spin-orbit resonance occurs, the maximum power dissipated inside the planets ranges between {{$10^{18}$ and $10^{19}$~W, while the total dissipated energy is of the order of $10^{30}-10^{32}$~J }} over a time interval as short as $(1-4.5) \times 10^{4}$~yr. }
   {Our illustrative models suggest that, if their host stars started their evolution on the ZAMS as fast rotators with periods between 0.5 and 1.0~days, the resonance occurred after about $40$~Myr since the host stars settled on the ZAMS in all the three cases. Such a huge heating over a so short time interval produces a complete melting of the planetary interiors and may shut off their hydromagnetic dynamos. These may initiate a successive phase of intense internal heating owing to unipolar magnetic star-planet interactions and affect the composition and the escape of their atmospheres, producing effects that could be observable during the entire lifetime of the planets. }


   \maketitle
%
%-------------------------------------------------------------------

\section{Introduction}

Planets with very short orbital periods have been discovered by observing their transits across the discs of their host stars. Considering those with an orbital period $P \la 1$ day, \citet{Sanchis-Ojedaetal14} found an occurrence rate ranging from $0.15 \pm 0.05$ percent for F-type dwarf stars to $1.1 \pm 0.4$ percent for M-type dwarfs with a prevalence of planets with radii between 0.8 and 1.25 Earth radii. For those planets, the occurence rate appears to be similar up to $P \la 2$~days \citepads{Fressinetal13,DressingCharbonneau15}. 

These planets are amenable to a measurement of their mass through radial velocity techniques because they induce  wobbles of their host stars with amplitudes of  a few m~s$^{-1}$ that are detectable with current spectrographs, while a correction of the stellar jitter due to magnetic activity can usually be performed \citepads[e.g.][]{Malavoltaetal18}. In some cases, they are also detectable in reflected light thanks to their closeness to their hosts, thus allowing a study of their atmospheric thermal emission and distribution of their surface brightness \citepads[see, e.g.,][for details]{Essacketal20}. Such studies are not presently possible for telluric planets orbiting at larger separations from their host stars.

Owing to the small distance from their host stars, such planets interact strongly with them through the gravitational, the radiation, the magnetic, and the stellar wind fields. Gravitation is responsible for very strong tidal interactions affecting the orbit and the rotation of these planets \citepads[e.g.,][]{Mathis18}, while the intense irradiation produces surface temperatures of thousands of kelvins on the illuminated hemisphere as well as a strong atmospheric evaporation powered by the stellar flux in the XUV spectral domain \citepads[][]{Owen19}. Magnetic interactions are also possible as recently discussed by, e.g., \citetads{Strugareketal19} in the case of \object{Kepler-78b}, a planet with an orbital period of only 8.5 hr. 

The close separation between these planets and their host stars leads to a synchronization of the planet rotation and a damping of the orbit eccentricity on timescales shorter than 1-10~Myr owing to the strong tidal interactions (see Sect.~\ref{model}). If the orbit eccentricity is not excited by other planets in the system, the internal heating of these planets will be dominated by the slow release of heat stored in the planet core after their formation and by the decay of radioactive elements \citep[e.g.,][]{Nimmoetal20}. However, another internal heating mechanism can operate during short phases of the planet evolution and this work is dedicated to introduce it. 

Young solar-like stars are usually fast rotators and show an intense magnetic activity sustained by a vigorous hydromagnetic dynamo. In these stars, strong magnetic fields with intensities up to several tens of teslas can be amplified and stored in the overshoot layers beneath their convection zones where such strong fields are probably organized in slender flux tubes that are stable against magnetic buoyancy thanks to the subadiabatic stratification of the layers. When their magnetic field intensity reaches a critical value, they become unstable to undulatory instability and emerge forming bipolar active regions in the photosphere characterized by the appearance of starspots \citepads[e.g.][and references therein]{Caligarietal95,Granzeretal00}. These strong-field flux tubes have a slightly lower density than the surrounding medium and are not distributed symmetrically in longitude over the star. Therefore, they produce a slight deviation of the stellar mass distribution from an axisymmetric configuration that in turn leads to a non-axisymmetric quadrupole component in the outer gravitational field of the rotating star (see Sect.~\ref{model}). 

When the stellar rotation period is twice the orbital period of the planet, such a quadrupole moment produces a resonant coupling between the stellar rotation and the orbital motion leading to an oscillation in the star-planet separation. As a consequence, a remarkable tidal dissipation is induced into the planet leading to an internal heating that can profoundly affect its internal structure  (see Sects.~\ref{model} and \ref{applications}). Although the spin-orbit resonance has a limited duration because of the evolution of the stellar rotation, the energy released inside the planet can produce a complete melting of its interior that may enhance its differentiation and affect the chemical composition of its surface and its atmosphere, thus leading to potentially permanent changes that could be observable (see Sect.~\ref{discussion}).

\section{Model}
\label{model}
In this section, we introduce the different ingredients of our model starting from the effects of strong magnetic flux tubes on the internal density stratification and the outer gravitational potential of the host star (Sect.~\ref{model_ft}); discussing the tidal deformation of the star and the planet (Sect.~\ref{tidal_def});  writing the Lagrangian function of the star-planet system from which we derive the equations of motion for its components (Sects~\ref{equations_of_motion} and~\ref{radial_motion}); and including the tidal damping of the orbital eccentricity (Sect.~\ref{planet_tidal_diss}).  Our equations are then solved  focussing on the radial orbital motion (Sects.~\ref{sol_rad_motion}) and the libration of the planet (Sects.~\ref{planet_libration} and~\ref{libration_power}). Finally, a simple model for the evolution of the stellar rotation and the orbital semimajor axis is introduced (Sect.~\ref{braking_and_tides}) in order to show how a resonance between the stellar rotation and the radial motion of the planet can happen along the evolution of the system. During such a resonance, a remarkable tidal dissipation occurs inside the planet as we shall see in some illustrative examples in Sect.~\ref{applications}. 

\subsection{The outer gravitational field of a magnetically active star}
\label{model_ft}
Late-type main-sequence stars have an internal structure characterized by an outer convection zone and a radiative interior, similar to the Sun. Immediately beneath their convection zone, there is a layer characterized by undershooting convection that is called the overshoot layer by analogy with a similar layer present in upper main-sequence stars \citep{Zahn91}. 
Such a layer is characterized by a subadiabatic stratification and a strong radial shear in the Sun \citep{Schouetal98} that allow to store and amplify intense magnetic fields. They reach  intensities between 10 and 200 T, depending on the depth of the convection zone and the rotation period of the star before becoming unstable and emerge through the convection zone. {Owing to the conservation of the angular momentum, the flux tubes do not emerge keeping their initial circular symmetry around the rotation axis {\citep[see, e.g.,][for a discussion of the force balance and the geometry of flux tubes initially in equilibrium]{Caligarietal95}}, but become non-axisymmetric with the azimuthal mode $m=1$ usually being the first to become unstable in rapidly rotating stars as differential rotation amplifies the field strength} \citep{vanBallegooijen82a,Moreno-Insertisetal92,Ferriz-MasSchussler94,Granzeretal00}. After emergence, these flux tubes are characterized by a strong buoyancy that makes them vertical with their roots anchored into the overshoot layer \citep{vanBallegooijen82b,Schussleretal94,Caligarietal95}. 

\begin{figure}
%\hspace*{-7mm}
 \centering{
 \includegraphics[width=9cm,height=7cm,angle=0,trim=87 87 93 93, clip]{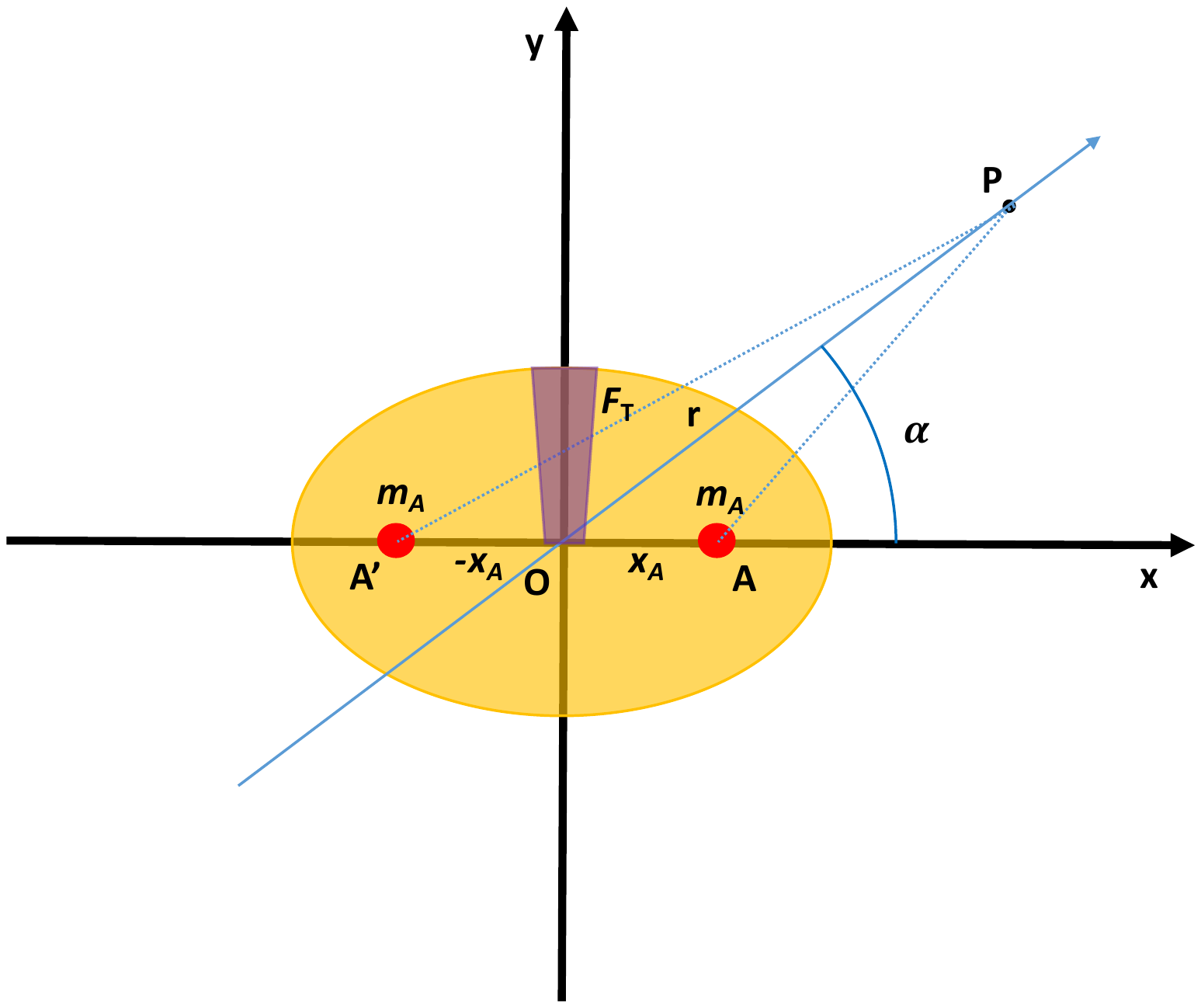}} % figure1.pdf}} % HD106252_1_rhkcorr_KR_final.pdf}}
% \vspace*{-10mm}
   \caption{Effect of the density perturbation inside a vertical magnetic flux tube $F_{\rm T}$ (indicated by the purple stub) in a stellar convection zone. For the purpose of evaluating the outer stellar gravitational potential, the mass deficit inside the tube is equivalent to adding two point masses $m_{\rm A}$ at the points $A$ and $A^{\prime}$ on a plane perpendicular to the tube axis here taken to be coincident with the $y$ axis.  The $x$-axis of our reference frame is taken along the line joining the two masses that are symmetrically situated with respect to the barycentre of the star $O$ at the abscissas $\pm \, x_{\rm A}$.  These two masses produce a gravitational potential outside the star that is no longer spherically symmetric and can be described by a quadrupole term. It depends on the distance $OP = r$ of a given point $P$ from the barycentre $O$ and the angle $\alpha$ between the direction $OP$ and  $OA$ of the line joining the two point masses (see the text).  }
              \label{vertical_flux_tube}%
\end{figure}

The density inside these vertical flux tubes is lower than in the surrounding medium because the external pressure is balanced in part by the interior magnetic pressure. {The equation governing the transversal mechanical equilibrium of a slender flux tube is \citep[e.g.,][and references therein]{vanBallegooijen82b}
\begin{equation}
p_{\rm i}(r) + \frac{B^{2}(r)}{2\mu} = p_{\rm e}(r),
\label{press_eq}
\end{equation}
where $p_{\rm i}(r)$ and $p_{\rm e}(r)$ are the plasma pressures inside and outside the flux tube, respectively, while $B(r)$ is the magnetic field strength at the radial distance $r$ from the centre of the star and $\mu$ the magnetic permeability of the plasma. By differentiating eq.~(\ref{press_eq}) with respect to the radius $r$ and using the equation for the radial pressure gradient $dp_{\rm i, \, e}/dr = - g(r) \,\rho_{\rm i, \, e} (r)$, we find
\begin{equation}
 \rho_{\rm i} (r) - \rho_{\rm e} (r) = \frac{1}{2\mu \, g(r)} \, \frac{dB^{2}(r)}{dr},
\label{dens_deficit}
\end{equation}
where $g(r)$ is the acceleration of gravity and $\rho_{\rm i, \, e}(r)$  the density of the plasma inside and outside the magnetic flux tube, respectively. Equation~(\ref{dens_deficit}) allows to compute the density deficit inside a slender magnetic flux tube when a model for the radial variation of the field inside the tube itself is specified. The simplest approach is to compute $B(r)$ by assuming that the field is mainly vertical due to the strong magnetic buoyancy and that $\nabla \cdot {\bf B}=0$. Such an approximation fails in the outermost layers of a star, but this has a very limited impact on the evaluation of the stellar quadrupole moment because those layers have a much lower density than the interior of the convection zone. } We refer the reader to  Sect.~2.3 of \citet{Lanza20a} for details on the calculations. A qualitative sketch of the effect of the density deficit in a vertical magnetic flux tube is given in Fig.~\ref{vertical_flux_tube} where the quadrupole gravitational potential associated with such a density deficit  is illustrated. 

The light curves of magnetically active stars and the Doppler imaging of their photospheres reveal that their magnetic flux tubes are unevenly distributed versus the stellar longitude \citep[see Sect.~2.1 of][for details]{Lanza20a}. {In addition to observations, hydromagnetic dynamo models also show the predominance of non-axisymmetric modes with an azimuthal wavenumber $m=1$ in stars rotating several times faster than the Sun \citep[e.g.,][]{Kapylaetal13,Vivianietal18,Kapyla20}}. Therefore, the net effect of several magnetic flux tubes simultaneously present in the convection zone of an active star is that of producing a non-axisymmetric gravitational quadrupole moment {when they are preferentially concentrated in one hemisphere as indicated by the observations and suggested by the above dynamo models. Sometimes,  configurations with two starspots of similar sizes on opposite hemispheres are observed \citep[e.g.,][]{CollierCameronetal09}, that point to an instability of magnetic flux tubes dominated by the $m=2$ mode as suggested by models in some specific regimes of stellar rotation and depth of the convection zone \citep[e.g.,][]{Granzeretal00}. In such a case, the gravitational quadrupole moment of the star is reduced, but the octupole moment could be significant and produce a similar resonance effect when the first harmonic of the rotation period corresponds to twice the orbital period of the planet. For simplicity, we shall not explore this possibility in the present work and limit ourselves to a model dominated by a non-axisymmetric density perturbation associated with the azimuthal $m=1$ mode. The single flux tube geometry depicted in Fig.~\ref{vertical_flux_tube} is adopted as the simplest configuration to model this case.  The effects of several flux tubes simultaneously present in the stellar convection zone can be computed by summing the contributions of the individual tubes because their effects on the quadrupole moment tensor are simply additive (see eqs.~\ref{qdef} and~\ref{inerdef} below). }

The outer gravitational potential $\Phi$ of the active star at a given point $P$ can be expressed as:
\begin{equation}
\Phi = -\frac{Gm_{\rm s}}{r} - \frac{3G}{2r^{3}} \sum_{i, k } \frac{Q_{ik} x_{i} x_{k}}{r^{2}},
\label{ogp}
\end{equation}
where $G$ is the constant of gravitation, $m_{\rm s}$ the mass of the star, $r$ the distance of $P$ from the barycentre $O$ of the star, $Q_{ik}$ the quadrupole moment tensor of the star, and $x_{i}$ the coordinates of $P$ in a Cartesian reference frame of origin $O$, while the indexes $i,  k = 1,2,3$ specify the Cartesian coordinates. %$x, y, z$, respectively. 
The quadrupole moment tensor can be expressed in terms of the inertia tensor of the mass distribution of the star as
\begin{equation}
Q_{ik} = I_{ik} - \frac{1}{3} \delta_{ik} {\rm Tr} {\vec I},
\label{qdef}
\end{equation}
where $\delta_{ik}$ is the Kronecker $\delta$ tensor and ${\rm Tr} {\vec I} = I_{xx} + I_{yy} + I_{zz}$ is the trace of the inertia tensor $\vec I$, that is, the sum of its diagonal components. The components of the inertia tensor are given by   
\begin{equation}
I_{ik} = \int_{V} \rho({\vec x}) x_{i} x_{k} \, dV,
\label{inerdef}
\end{equation}
where $\rho$ is the density, $\vec x$ the position vector, and $V$ the volume of the star over which the integration is extended.

Equation~(\ref{ogp}) can be simplified by adopting a Cartesian reference frame whose axes  are directed along the principal axes of inertia of the star so that only the diagonal components of the inertia tensor are different from zero. The principal axes of inertia are generally coincident with the axes of symmetry of the star. We assume that the $z$-axis is directed along its rotation axis, while the $x$- and $y$-axes are in the equatorial plane of the star. Introducing a set of spherical polar coordinates $(r, \theta, \alpha )$, where $r$ is the distance from the barycentre $O$ of the star, $\theta$ the colatitude measured from the $z$-axis, and $\alpha$ the longitude, we have:
\begin{equation}
\Phi (r, \theta, \alpha) = -\frac{Gm_{\rm s}}{r} -\frac{3G}{2 r^{3}} \left( -2 Q + 3Q \sin^{2} \theta + \frac{1}{2} T \sin^{2} \theta \cos 2\alpha \right), 
\label{star_pot}
\end{equation}
where the two scalars $Q \equiv (Q_{xx} + Q_{yy})/2 = - Q_{zz}/2$ and $T \equiv Q_{xx} - Q_{yy}$ express the axisymmetric and non-axisymmetric components of the quadrupole tensor $Q_{ik}$ referred to its principal axes, respectively. In that reference frame, only two scalars are required to specify the tensor  because it is traceless by definition (cf. ~eq.~\ref{qdef}). 

Assuming that the magnetic flux tube in Fig.~\ref{vertical_flux_tube} lies in the equatorial plane of the star along the $y$-axis, the principal axes of inertia will be the $y$-axis itself and the two perpendicular axes $x$ in the equatorial plane and $z$ orthogonal to the equatorial plane. By applying eq.~(\ref{inerdef}), we obtain $I_{xx} = 2 m_{\rm A} x_{\rm A}^{2}$, $I_{yy}= I_{zz}=0$, and ${\rm Tr} {\vec I } = I_{xx}$, where $m_{\rm A}$ and $x_{\rm A}$ are the mass and the distance from the barycentre of each of the two point masses $A$ and $A^{\prime}$ that simulate the effect of the density deficit inside the magnetic flux tube. In this way, we find $Q = (1/3)\, m_{\rm A} x_{\rm A}^{2}$ and $T = 2\, m_{\rm A} x_{\rm A}^{2}$. Substituting these expressions into eq.~(\ref{star_pot}), we obtain the outer gravitational potential of the star in this simple approximation. 

The quadrupole term coincides with the expression obtained by developing the potentials of the individual point masses $m_{\rm A}$ at the outer point $P$ in series of the small ratio $(|x_{\rm A}|/r) \ll 1$ up to the second order \citep[cf.][Sect.~5.3]{MurrayDermott99}. It shows that the principal axis of inertia $x$ coincides with the line joining the two point masses that is also the axis of symmetry of the bulge of the density distribution of the star. Note that the moment of inertia for rotation about the $x$-axis, that is, $I_{yy} + I_{zz}$,  is minimal. 

Going beyond this simplified model, the quadrupole terms produced by a vertical flux tube in the equatorial plane have been computed by \citet{Lanza20a} {following a method based on eqs.~(\ref{press_eq}) and~(\ref{dens_deficit}) \citep[see Sects.~2.3 and 2.4 of][for details]{Lanza20a}}. Considering the configuration depicted in Fig.~\ref{vertical_flux_tube} with the axis of the flux tube directed along the $y$-axis of the reference frame and the angle $\alpha$ measured from the $x$-axis, they are 
\begin{eqnarray}
Q = \frac{\pi}{3} \frac{B_{0}^{2} r_{\rm b}^{4}}{\mu} {\cal J} \sin^{2} \theta_{0} \cos \theta_{0}, \label{qandt0} \\
T = 2\pi \frac{B_{0}^{2} r_{\rm b}^{4}}{\mu} {\cal J} \sin^{2} \theta_{0} \cos \theta_{0},
\label{qandt}
\end{eqnarray}
where $B_{0}$ is the intensity of the magnetic field at the base of the flux tube in the overshoot layer that is at a radial distance $r_{\rm b}$ from the centre of the star, $\theta_{0}$ the angular radius of the flux tube, and $\cal J$ is the integral
\begin{equation}
{\cal J} = \int_{r_{\rm b}}^{r_{\rm L}} \frac{r^{\prime}}{G M(r^{\prime})} \, dr^{\prime},
\label{j_integ}
\end{equation}
where  $M(r)$ is the mass of the star inside the radius $r$ and $r_{\rm L}$ the limit radius beyond which the magnetic field of the flux tube deviates significantly from the vertical because the pressure of the plasma outside the flux tube becomes too small to confine the field. The exact value of $r_{\rm L}$ is not critical because the density in the outer layers of the star where the field is no longer confined by the plasma pressure is so small that the effect on the quadrupole moment is negligible \citep[cf.][]{Lanza20a}. 
If the fraction of the stellar surface covered by the section of the magnetic flux tube is $f_{\rm s}$, the angle $\theta_{0}$ is given by $\theta_{0} =\arccos (1-2f_{\rm s})$ or $\theta_{0} \simeq 2 \sqrt{f_{\rm s}}$ for $f_{\rm s} \ll 1$. 

In \citet{Lanza20a}, the axis of the flux tube was directed along the $x$-axis of the reference frame in Fig.~\ref{vertical_flux_tube}, thus there was a minus sign in the r.h.s. of eqs.~(\ref{qandt0}) and~(\ref{qandt}). Choosing to have the tube axis along the $y$-axis instead of along the $x$-axis is equivalent to add $\pi/2$ to the angle $\alpha$ in eq.~(\ref{star_pot}) which compensates for the sign change of $Q$ and $T$ with respect to the expressions in \citet{Lanza20a}. The present choice appears to be more motivated from a physical point of view because the density deficit inside the flux tube corresponds to adding two excess masses $m_{\rm A}$ along the $x$-axis perpendicular to the tube axis. With  the simple two-mass model, we obtain the above quadrupole moment components by assuming that the effective moment of inertia $m_{\rm A} x_{\rm A}^{2}$ is equal to $\pi \, (B_{0}^{2} r_{\rm b}^{4}/\mu)\, {\cal J}  \sin^{2} \theta_{0} \cos \theta_{0}$. 
%%%%%%%%%%%%%%%%%%%%%%%%%%%%%%%%%%%%%
%%%%%%%%%%%%%%%%%

\subsection{Origin of the stellar and planet quadrupole moments}
\label{tidal_def}
The effect of the tidal potentials acting on the star and the planet is that of producing a deviation of their density distributions from axial symmetry. The outer gravitational potential of each body can be developed in a series of spherical harmonics the angular dependence of which is given by the Legendre polynomials $Y_{lm}$ \citep[e.g.,][]{Ogilvie14}. The term of the tidal potential of degree $l$ and azimuthal order $m$ produces a density perturbation in each body that is responsible for a component of the outer gravitational potential with the same degree and azimuthal order, but with a different amplitude that depends on the internal stratification and the rigidity of the body. 

The ratio of the amplitude of the outer gravitational component of degree $l$ and order $m$ to the amplitude of the corresponding component of the tidal potential is indicated with $k_{lm}$ and is called the Love number of degree $l$ and azimuthal order $m$ of the body. It is in general a complex number the imaginary part of which expresses the tidal lag and depends on the frequency of oscillation $\hat{\omega}$ of the tidal potential.  The real part of $k_{lm}$ expresses the hydrostatic deformation of the body under the action of the tidal potential and is generally much larger than its imaginary part because the tidal lag is usually a very small angle, especially in the case of fluid bodies such as a star. Since the most important component of the tidal potential corresponds to $l=m=2$, we shall focus on the Love number $k_{22} = k_{2}$, that is the second-order Love number. It can be used to express the outer quadrupole potential of a body produced by the equilibrium tide. It  corresponds to an ellipsoidal deformation whose axis of symmetry is always directed along the line joining the barycentres of the two bodies (the perturbed one and the companion producing the tidal potential), if we neglect the very small tidal lag angle. 

In the case of a fluid body, like a star or a giant planet, $k_{2}$ is entirely defined by the function $\rho(r)$ that specifies the unperturbed density of the body as a function of the distance $r$ from its barycentre and that is spherically symmetric \citep[e.g.,][]{Kopal59}. However, in the case of a telluric planet the value of $k_{2}$ depends also on the rigidity of the body that takes into account the deviation of its response from the purely hydrostatic case. Therefore, the quadrupole moment of a planet of mass $m_{\rm p}$ and with a radius $R_{\rm p}$ at a separation $D$ from a star of mass $m_{\rm s}$ can be expressed as \citep[cf.][]{Remusetal12}
\begin{equation}
T = k_{2} \left( \frac{m_{\rm s}}{m_{\rm p}} \right) \left( \frac{R_{\rm p}}{D} \right)^{3} m_{\rm p} R_{\rm p}^{2};   
\label{remus_teq}
\end{equation}
%In the case of a giant planet, $k_{2} \sim 0.36$ because its internal stratification is close to that of an isothermal polytrope, while 
for the Earth $k_{2} = 0.295$ \citep{Lainey16}, while for a fluid body of uniform density $k_{2} = 3/2$ \citep{Ogilvie14}. 

In the case of the host star in a planetary system, the hydrostatic deformation due to the tides raised by the planet produces an {\em hydrostatic} quadrupole moment that can be neglected for our purposes because the tidal bulge is always directed along the line joining the centres of the two bodies. The tidal lag angle $\alpha_{\rm T}$, that measures the deviation of the axis of symmetry of the bulge from the line joining the barycentres of the two bodies, is negligible. More precisely, it is given by $\alpha_{\rm T} \simeq 1/Q^{\rm T}_{\rm s}$, where $Q_{\rm s}^{T}$ is the tidal quality factor of the star \citep{Ogilvie14} \footnote{We use the symbol $Q^{\rm T}$ to avoid confusion with the quadrupole term $Q$ introduced in Sect.~\ref{model_ft}.}, that is of the order of at least $10^{4}-10^{6}$ in the case of late-type stars \citep[][see also below Sect.~\ref{braking_and_tides}]{OgilvieLin07}.

The situation is different in the case of the non-axisymmetric quadrupole produced by the density perturbation inside the magnetic flux tubes in the stellar convection zone that we considered in Sect.~\ref{model_ft}. In that case, the bulge of the density perturbation is not constrained to stay close to the line joining the barycentres of the two bodies, so the angle $\alpha$ in Fig.~\ref{vertical_flux_tube} can assume any value. 

The magnetic flux tubes we are considering are not permanent structures because their magnetic fields are subject to turbulent diffusion, thus they have a finite lifetime in stellar convection zones. In the case of the Sun, the longest lived magnetic flux tubes are associated with recurrent active regions that last a few solar rotations, that is, a few months { \citep[e.g.,][]{Solanki03,Hathaway08}}. Nevertheless, in the case of young active stars, active regions can persist for several years at the same active longitude because they are much larger and the diffusion timescale is proportional to the area of the region. Lifetimes of hundreds of days are observed in Sun-like stars with rotation periods $P_{\rm rot}$ shorter than $5-7$ days  \citep[e.g.,][]{Gilesetal17}, while  active longitudes persist for decades in rapidly rotating stars ($P_{\rm rot} \la 2-4$~days) with deep convection zones showing a very strong dynamo action \citep[e.g.,][]{Lanzaetal06}. Therefore, we can regard these active stars as endowed with non-axisymmetric quadrupole moments that persist for timescales ranging from hundreds of days to several years.  Only the dynamical effect of this component of the non-axisymmetric quadrupole moment will be considered in our model because the value of the angle $\alpha$ can change in time, while we shall neglect the quadrupole associated with the equilibrium tides because $\alpha_{\rm T}$ is extremely small and thus gives a virtually constant contribution to the potential that can be neglected for our purposes (cf. Sect.~\ref{equations_of_motion}).  

Finally, we consider that a telluric planet may have a permanent quadrupole moment associated with its solid layers in addition to the tidal (hydrostatic) quadrupole deformation. In the case of the Earth, the hydrostatic deformation is prevalent and $Q^{\rm T}_{\rm p} \sim 280$ considering the Earth-Moon tidal interaction \citep{Lainey16}; this implies that $\alpha_{\rm T}$ is always small {(see Sect.~\ref{planet_tidal_diss} for more information on $Q^{\rm T}_{\rm p}$ in the case of telluric planets)}. Conversely, in the case of the Moon or Mercury, the permanent quadrupole moment associated with the persistent deviation of their bodies from axisymmetry is prevalent and is responsible for the spin-orbit coupling experienced by those bodies \citep[][Ch.~5]{GoldreichPeale68,MurrayDermott99}. Therefore, we shall include the contribution of a possible permanent quadrupole moment of the planet in our model because the corresponding lag angle $\alpha$ is not a priori constrained to remain small. 

\subsection{Lagrangian function and equations of motion of the star-planet system}
\label{equations_of_motion}

To study the dynamics of our star-planet system, we apply the Lagrangian formalism \citep[e.g.,][]{Goldstein50}. The Lagrangian $\cal L$ of our system is defined as
\begin{equation}
{\cal L} = {\cal T} - \Psi,
\end{equation}
where $\cal T$ is the kinetic energy of the system and $\Psi$ its potential energy expressed as functions of the coordinates and their time derivatives in an inertial reference frame. 
We consider a reference frame having its origin at the barycentre $Z$ of the star-planet system and the $xy$ plane of which is the orbital plane of the system. In this reference frame, the kinetic energy of the system can be written as the sum of the kinetic energy of the relative orbital motion and of the kinetic energy of rotation of each of the bodies around an axis passing through its barycentre, respectively. For the sake of simplicity, we assume that the spin of the star and of the planet are perpendicular to the orbital plane. Tides inside the planet induce the alignment and the synchronization of its spin with the orbital motion on a timescale of the order of $\la 10^{5}$ yr, much shorter than the lifetime of the system \citep[][]{Leconteetal10}. Therefore, the only assumption that may not be always verified is that the star spin has zero obliquity. Nevertheless, we make this assumption to keep our model as simple as possible, noting that there are some counterexamples to it \citep[see][for more details and possible interpretations]{Beckeretal20}. 

\begin{figure}
%\hspace*{-7mm}
 \centering{
 \includegraphics[width=9cm,height=6cm,angle=0,trim=87 87 93 93, clip]{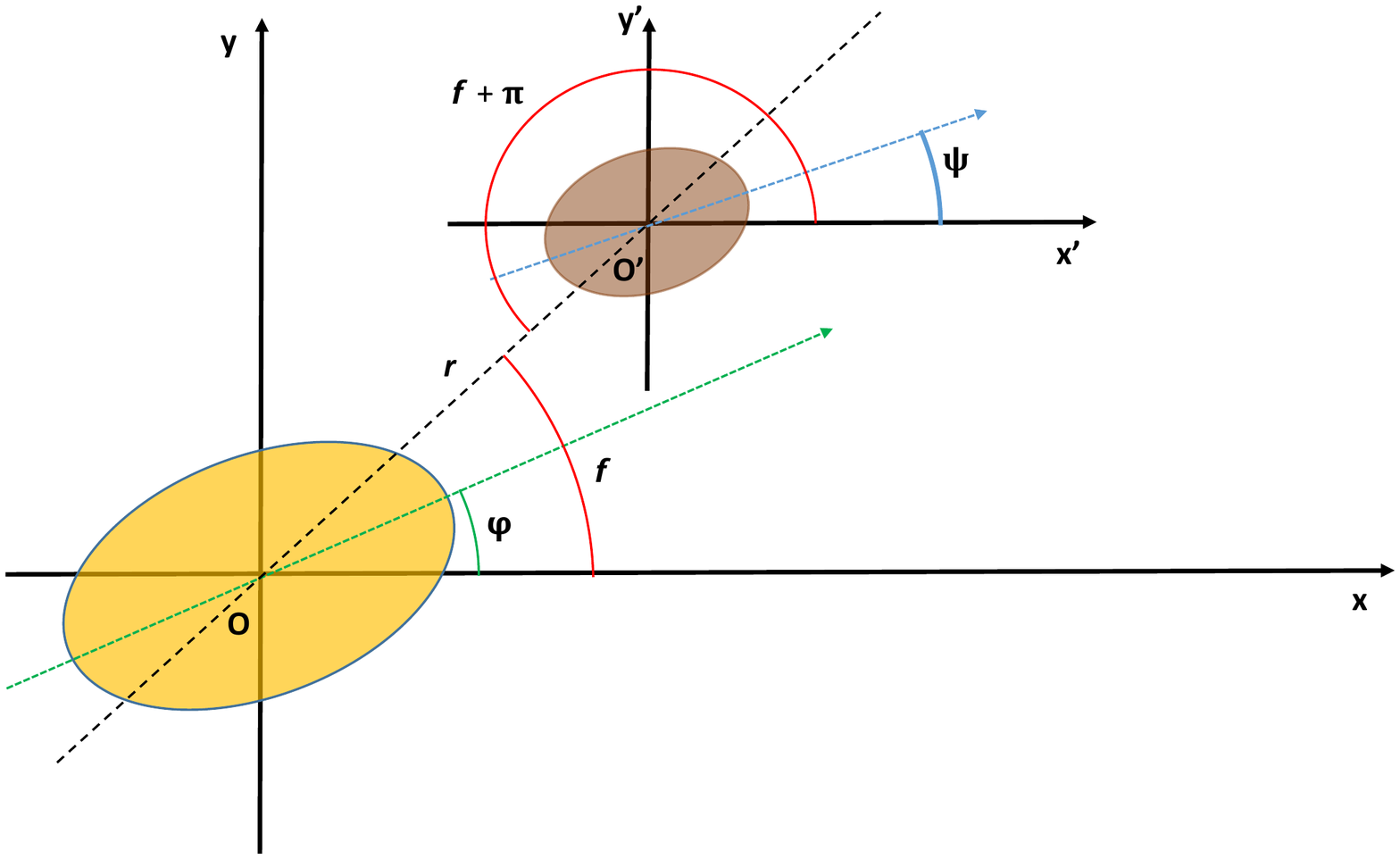}} % figure2.pdf}} % HD106252_1_rhkcorr_KR_final.pdf}}
% \vspace*{-10mm}
   \caption{Reference frames adopted to define the true anomaly $f$ and the rotational coordinates of the star $\varphi$ and the planet $\psi$ to express the Lagrangian of our star-planet system (see text for details).   The relative separation of the barycentres of the two bodies is $r = OO^{\prime}$. The rotational coordinate of the star is defined as the angle between the $x$-axis, the direction of which is fixed in an inertial space, and the principal axis of inertia of the star with respect to which the angle $\alpha$ in eq.~(\ref{star_pot}) is measured. A similar definition is adopted for the rotational coordinate of the planet considering $\psi$ as the angle between the axis $x^{\prime}$ parallel to $x$ and the  principal axis of inertia of the planet passing through its permanent quadrupole bulge (see text). }
              \label{ref_frame}%
\end{figure}

The kinetic energy of the orbital motion of our system can be expressed as the energy of the relative motion of a body having the reduced mass of the system $m = m_{\rm s} m_{\rm p}/(m_{\rm s}+ m_{\rm p})$, where $m_{\rm s}$ is the mass of the star and $m_{\rm p}$ that of the planet, around the barycentre $Z$ of the system. The kinetic energy of rotation of each body can be expressed in terms of the moment of inertia around their spin axes and the angular velocity of rotation around those axes. Specifically, we write:
\begin{equation}
{\cal T} = \frac{1}{2} m \left( \dot{r}^{2} + r^{2} \dot{f}^{2} \right) + \frac{1}{2} I_{\rm s} \dot{\varphi}^{2} + \frac{1}{2} I_{\rm p} \dot{\psi}^{2},
\label{kin_ener}
\end{equation}
where $r$ is the relative distance between the two bodies, $f$ the true anomaly of their relative orbital motion, $I_{\rm s}$ the moment of inertia of the star around its rotation axis, $\varphi$ the rotational coordinate of the star, $I_{\rm p}$ the moment of inertia of the planet around its rotation axis, and $\psi$ the rotational coordinate of the planet; the dot over a variable indicates its time derivative.  The angles $f$, $\varphi$, and $\psi$ are illustrated in Fig.~\ref{ref_frame}, where the barycentres of the star and the planet are $O$ and $O^{\prime}$, respectively, and the $xy$ and $x^{\prime}y^{\prime}$ planes coincide with the orbital plane. The Cartesian reference frames $xy$ and $x^{\prime} y^{\prime}$ have their origins at $O$ and $O^{\prime}$, respectively, and their axes have fixed directions in an inertial space. Specifically, the angles $\varphi$ and $\psi$ are measured from the $x$ and $x^{\prime}$ axes to the principal axes of inertia corresponding to the quadrupole bulges of the star and the planet, respectively (see Sect.~\ref{model_ft}). This definition simplifies the expression of the gravitational potential energy $\Psi$  as
%%%%%%%%%%%%%%%%%%%%%%%%%%%%%%%%%%%%%
%%%%%%%%%%%%%%%%%
\begin{eqnarray}
\Psi  & = & -\frac{G m_{\rm s} m_{\rm p}}{r} - \frac{3Gm_{\rm p}}{2 r ^{3}} \left[ Q_{\rm s} +\frac{1}{2} T_{\rm s} \cos 2 (f-\varphi) \right] + \nonumber \\
& &  - \frac{3Gm_{\rm s}}{2 r ^{3}} \left[ Q_{\rm p} +\frac{1}{2} T_{\rm p} \cos 2 (f + \pi -\psi) \right],
\label{pot_ener}
\end{eqnarray}
where $Q_{\rm s}$ and $T_{\rm s}$ are the axisymmetric and non-axisymmetric quadrupole moments of the star as defined in Sect.~\ref{model_ft}, while $Q_{\rm p}$ and $T_{\rm p}$ are the same quantities for the planet (see below). 

The potential energy of the system in eq.~(\ref{pot_ener}) includes the first-order term of the gravitational energy that assumes that the masses of the two bodies are concentrated into their barycentres and the second-order term associated with their gravitational quadrupole moments with the stellar quadrupole acting on the planet and the planet quadrupole acting on the star, respectively. The quadrupole terms of the potential energy follow from the expression~(\ref{star_pot}) of the quadrupole potential written for the case of the equatorial plane of the star and the planet ($\theta = \pi/2$) and considering that the angle $\alpha = f -\varphi$ in the case of the star or $\alpha=f+\pi - \psi$ in the case of the planet (cf. Fig.~\ref{ref_frame}). %thanks to the definition of the rotational coordinates $\varphi$ and $\psi$ that are defined by means of  the principal axes of inertia of the two bodies, respectively (cf. Fig.~\ref{ref_frame}). 

The bulges due to the equilibrium tides are almost aligned with the line joining the barycentres $O$ and $O^{\prime}$ of the two bodies and their magnitudes are almost constant for a nearly circular orbit. Therefore, they can be assumed to contribute to $Q_{\rm s, p}$, respectively, while only the persistent non-axisymmetric quadrupole of magnetic origin is included into $T_{\rm s}$ and only the permanent deformation of the planet gives rise to $T_{\rm p}$ in our model (see the discussion in Sect.~\ref{tidal_def}). 

The equations of motion of our system can be derived from its Lagrangian as
%\begin{widetext}
%\hline
\begin{equation}
\frac{d}{dt} \left( \frac{\partial {\cal L}}{\partial \dot{q}}  \right) - \frac{\partial {\cal L}}{\partial q} = 0,
\end{equation}
where $t$ is the time and $q = r, f, \varphi,$ or $\psi$ is any of the coordinates adopted to describe the motion of the system. In this way, we obtain the following equations of motion:
\begin{eqnarray}
 m \ddot{r} -mr \dot{f}^{2}   +  \frac{Gm_{\rm s} m_{\rm p}}{r^{2}}  
 +  \frac{9G}{2 r^{4}} \left[ Q_{\rm s} m_{\rm p}+ \frac{1}{2} T_{\rm s}  m_{\rm p} \cos 2 (f-\varphi)  + \right. \nonumber   \\ 
\left. + Q_{\rm p} m_{\rm s}+ \frac{1}{2} T_{\rm p} m_{\rm s} \cos 2 (f-\psi) \right]  =  0,   \label{r_eq} \\
 m \frac{d}{dt} \left( r^{2} \dot{f} \right) + \frac{3G}{2r^{3}} \left[ T_{\rm s} m_{\rm p} \sin 2 (f-\varphi)  +  \right.  \nonumber \\ 
\left. + T_{\rm p} m_{\rm s} \sin 2( f-\psi) \right] = 0, \label{f_eq} \\
 I_{\rm s} \ddot{\varphi} -\frac{3G}{2 r^{3}} T_{\rm s} m_{\rm p} \sin 2(f-\varphi)  =  0,  \label{phi_eq} \\
 I_{\rm p} \ddot{\psi} -\frac{3G}{2 r^{3}} T_{\rm p} m_{\rm s} \sin 2(f-\psi)  =   0. \label{psi_eq} 
\end{eqnarray}
%\hline
%\end{widetext}
Summing equations~(\ref{f_eq}), (\ref{phi_eq}), and~(\ref{psi_eq}) and integrating with respect to the time, we obtain an equation expressing the conservation of the total angular momentum $J$ of the system:
\begin{equation}
m r^{2} \dot{f} + I_{\rm s} \dot{\varphi} + I_{\rm p} \dot{\psi} = J.  
\end{equation}
The spin angular momentum of the planet is usually negligible because its rotation is synchronized by tides ($\dot{f} \sim \dot{\psi}$) and $I_{\rm p} \ll mr^{2}$ for $m_{\rm p} \ll m_{\rm s}$. The angle $f-\psi$ is always very small because %the hydrostatic tidal bulge has always a minuscule deviation from the line joining the barycentres of the star and the planet when the planet is mainly fluid, while 
it makes damped oscillations about the equilibrium zero value  as we shall see from eq.~(\ref{beta_eq_com}) below.
Therefore, we can simplify eq.~(\ref{r_eq}) by  considering that when the planet spin is synchronized and the orbit is close to circular, $\cos 2(f -\psi) \simeq 1 $ to the first order,  thus giving
\begin{equation}
\ddot{r} - r \dot{f}^{2} + \frac{Gm_{\rm T}}{r^{2}} + \frac{9Gm_{\rm T}}{2 m_{\rm s} r^{4}} \left[\tilde{Q}_{\rm s} + \frac{1}{2} T_{\rm s} \cos 2 (f- \varphi) \right] = 0, 
\label{rnew_eq}
\end{equation}
where $m_{\rm T} \equiv m_{\rm s} + m_{\rm p}$ is the total mass of the system and we have defined
\begin{equation}
\tilde{Q}_{\rm s} \equiv Q_{\rm s} + \left( \frac{m_{\rm s}}{m_{\rm p}} \right) \left( Q_{\rm p} + \frac{1}{2} T_{\rm p} \right). 
\end{equation}

\subsection{Radial motion}
\label{radial_motion}

The eccentricity of the orbit can be assumed to be small because it is rapidly damped by the tides inside the planet (see Sect.~\ref{planet_tidal_diss}). Therefore, the terms involving the eccentricity will be of the first order as well as all the terms involving the quadrupoles, while their products will be of the second order, so we can neglect them. Since the angle $f -\psi \sim 0$, we neglect the contribution of the planet spin to the variation of the orbital angular momentum and  re-write eq.~(\ref{f_eq}) as
\begin{equation}
m \frac{d}{dt} \left( r^{2} \dot{f} \right)  \simeq - \frac{3G}{2a^{3}}  T_{\rm s} m_{\rm p} \sin  [2(n-\Omega)t], 
\label{f_eq1}
\end{equation}
where $a$ is the orbit semimajor axis; $n$ the orbital mean motion, $n \equiv 2\pi/P_{\rm orb}$, with $P_{\rm orb}$  the orbital period; and $\Omega \equiv 2\pi/P_{\rm rot}$ the spin angular velocity of the star with $P_{\rm rot}$ its rotation period.  Equation~(\ref{f_eq1}) is a first-order equation valid for $e \ll 1$ and can be immediately integrated with respect to the time to yield for $n \not= \Omega$
\begin{equation}
\dot{f} = n \left( \frac{a}{r} \right)^{2}  \left\{ 1 + \epsilon \cos \left[2 (n -\Omega) t \right] \right\},
\label{fdot_eqn}
\end{equation}
where 
\begin{equation}
\epsilon \equiv \frac{3}{4} \left( \frac{T_{\rm s}}{m_{\rm s} a^{2}} \right) \left( \frac{n}{n-\Omega} \right),
\end{equation}
is a very small quantity ($\epsilon \ll 1$) because the non-axisymmetric component of the stellar quadrupole moment $T_{\rm s}$ is very small in comparison with the moment of inertia of the orbit $m a^{2}$ (see Sect.~\ref{applications}). %In general, all the terms involving the stellar quadrupole moments are first order in comparison the those related to the orbital motion in the absence of quadrupole perturbations. 
When the orbit is nearly circular, we can define the small non-dimensional quantity $x(t)$ such that $ |x(t) | \ll 1$ from 
\begin{equation}
r \equiv a(1+x),
\label{x_defin}
\end{equation}
and substitute it into eq.~(\ref{rnew_eq}) taking only the first-order terms. In this way, making use also of eq.~(\ref{fdot_eqn}), we obtain an equation for $x$ for $n \not= \Omega$:
\begin{equation}
\ddot{x} + n^{2} \left\{ 1 + \eta + \zeta \cos \left[2 (n - \Omega) t \right]\right\} x = \xi + f(t), 
\label{x_eq}
\end{equation}
where $\eta$, $\zeta$, and $\xi$ are small quantities defined as
\begin{eqnarray}
\eta \equiv - 18 \frac{\tilde{Q}_{\rm s}}{m_{\rm s} a^{2}}, \\
\zeta \equiv 9 \left( \frac{T_{\rm s}}{m_{\rm s}a^{2}} \right) \left( \frac{2\Omega-n}{2n - 2\Omega} \right), \; \mbox{and} \\ 
\xi \equiv - \frac{9}{2} n^{2} \left( \frac{\tilde{Q}_{\rm s}}{m_{\rm s} a^{2}} \right) = \frac{1}{4} n^{2} \eta,
\end{eqnarray}
while the forcing term 
\begin{equation}
f(t) \equiv A_{\rm x} \left[ \frac{T_{\rm s}(t)}{I_{\rm s}} \right] \cos \left[2(n-\Omega) t \right]
\label{fforcing}
\end{equation}
has an amplitude $A_{\rm x}$ given by
\begin{equation}
A_{\rm x} \equiv \frac{3}{2} n ^{2} \left( \frac{I_{\rm s}}{m_{\rm s} a^{2}} \right) \left( \frac{3\Omega - n}{2n -2\Omega} \right). 
\label{fforcing_amp}
\end{equation}
In eq.~(\ref{fforcing}), we indicated explicitly the dependence of the stellar quadrupole term $T_{\rm s}$ on the time because of the finite lifetime of the magnetic flux tubes that are responsible for the term itself and normalized it to the stellar moment of inertia $I_{\rm s}$ instead of the moment of inertia $m_{\rm s} a^{2}$ that explains the additional factor appearing in eq.~(\ref{fforcing_amp}). 

\subsection{Tidal damping of the eccentricity of the planetary orbit}
\label{planet_tidal_diss}

Equation~(\ref{x_eq}) implies that $x$ can make free oscillations along the orbit with the same frequency $n$ of the orbital motion because $\eta, \zeta \ll 1$. Such oscillations correspond to the motion along an instantaneously eccentric orbit, therefore dissipative tides are excited within the planet and the star. They lead to the damping of the oscillations themselves with a timescale $\tau_{\rm e} \equiv -e/(de/dt) $  that is given by \citep[cf.][]{Jacksonetal08}:
\begin{equation}
\tau_{\rm e}^{-1} = \left[ \frac{63 R_{\rm p}^{5} \sqrt{G m_{\rm s}^{3}}}{4 Q^{\prime \rm T}_{\rm p} m_{\rm p}} + \frac{225 R_{\rm s}^{5} m_{\rm p} \sqrt{G/m_{\rm s}}}{16 Q^{\prime \rm T}_{\rm s}}\right] a^{-13/2},
\label{tau_ecc}
\end{equation}
where $R_{\rm s}$ and $R_{\rm p}$ are the radii of the star and the planet, $m_{\rm s}$ and $m_{\rm p}$ their masses, respectively, while $Q^{\prime \rm T}_{\rm s}$ and $Q^{\prime \rm T}_{\rm p}$ are the modified tidal quality factors of the star and the planet defined as $Q^{\prime \rm T}_{\rm s, p} \equiv (3/2) (Q_{\rm s, p}^{\rm T} / k_{2 \, \rm s, p})$ \citep[see][]{Ogilvie14}. {Note that the smaller the tidal quality factor, the larger the  dissipation of the tidal kinetic energy and the damping of the eccentricity. 

The tidal dissipation inside a telluric planet depends on the viscoelastic properties of its interior that in turn depends on its structure, composition, and the frequency of the tidal forcing. In our case, the tidal  frequency corresponds to the orbital mean motion because we consider the eccentricity tide excited by the varying separation of the planet from its host star along its orbit. The simplest treatment, based on the Maxwell rheological model, severely underestimates the dissipation for tidal frequencies much higher than the inverse of the so-called Maxwell time, which is of the order of $\sim 10^{4}$~yr for the Earth. Conversely, the Andrade rheological model gives a better description of the dissipation at the high frequencies characteristic of our case \citep{Tobieetal19,Bolmontetal20}. The multi-layered structure of a telluric planet also plays a relevant role, especially if it has an inner liquid core, a differentiated mantle, or a water ice layer, thus  making simple models based on an homogeneous structure usually overestimating the  dissipation \citep[cf. Fig.~6 in][]{Bolmontetal20}. 

The values of $Q^{\rm T}_{\rm p}$ and the Love numbers $k_{2}$ computed by \cite{Tobieetal19} for telluric planets consisting of a liquid metallic core and a silicate mantle, divided into an upper and a lower layer to account for a mineralogical transition occurring at pressures around $25$~GPa, can be adopted for our investigation. Their model reproduces the observed tidal quality factors and Love numbers of the Earth at different tidal frequencies. Considering a forcing period of $1$~day, their computed $Q^{\rm T}_{\rm p}$ and $k_{2}$ can be applied to the case of our telluric planets with orbital periods between $4$ and $24$ hours because their dependencies  on the tidal frequency in this range are smaller than the effect of the unknown interior structure and composition \citep[e.g., Fig.~6 of][]{Tobieetal19}. The value of $Q^{\prime \rm T}_{\rm p} = (3/2) (Q^{\rm T}_{\rm p}/k_{2})$ increases from $\sim 10^{3}$ for a planet with the mass of the Earth, up to $\sim 2 \times 10^{3}$ for a super-Earth of $8-10$ Earth masses. Comparable results have been obtained by \citet{Bolmontetal20}\footnote{\citet{Bolmontetal20} use the imaginary part $\Im k_{2}$ of the Love number $k_{2}$ to express the tidal dissipation which is related to $Q^{\prime \rm T}_{\rm p}$ by $\Im k_{2} = 3/(2Q^{\prime \rm T}_{\rm p})$. }. Therefore, we can adopt the measured value of $Q^{\prime \rm T}_{\rm p} = 1425$, relative to the semi-diurnal tide of our Earth with a forcing period of $12.4$ hours \citep{Lainey16}, as representative also of our telluric planets, although they are not expected to have surface oceans as our Earth. {In other words, the above $Q^{\prime \rm T}_{\rm p}$ value is assumed to refer to the mantle only, while in the present Earth the $Q^{\prime \rm T}_{\rm p}$ value associated with the oceanic tides is smaller and the tidal dissipation is actually dominated by the friction  applied by the topography on tidal gravito-inertial waves that propagate in the oceans \citep[e.g.,][]{EgbertRay00,EgbertRay03,Tyler21}.}

The intense tidal heating experienced by a telluric planet as we shall find in our model, can deeply modify its rheological properties. For example, the melting of its silicate mantle can produce a decrease of its viscosity by several orders of magnitude that in turn can increase tidal dissipation \citep[e.g., ][]{Walterova20}. As we shall see, these complications as well as the ignorance of the interior structure and mineralogical composition of ultra-short period planets, will not affect our conclusions  because the total dissipated energy inside the planet will not depend critically on the value of $Q^{\prime \rm T}_{\rm p}$. Therefore, we shall assume the above Earth value for reference in our investigation. } 

In our case, the tidal dissipation inside a telluric planet dominates that inside the star because $Q^{\prime \rm T}_{\rm p} \ll Q_{\rm s}^{\prime \rm T}$, {given  that $Q^{\prime \rm T}_{\rm s} \ga 10^{5}-10^{6}$ as we shall see in Sect.~\ref{braking_and_tides}}. As a consequence, we shall assume that tidal dissipation is dominated by tides inside the planet and that all the dissipated mechanical energy is converted into heat inside the planet.   

In conclusion, we can recast equation~(\ref{x_eq}) including the damping effects of tides as
\begin{equation}
\ddot{x} + 2 b \dot{x} + n^{2} \left\{ 1 + \eta + \zeta \cos \left[2 (n - \Omega) t \right]\right\} x = \xi + f(t),
\label{x_eq1}
\end{equation}
where $2b \equiv \tau_{\rm e}^{-1}$. By multiplying both sides of eq.~(\ref{x_eq1}) by $m a^{2} \dot{x}$, we obtain
\begin{equation}
\frac{d {\cal E}_{\rm R}}{dt} \simeq -2bm\,a^{2} \dot{x}^{2} + m a^{2} \dot{x} \left[ f(t) + \xi \right],
\label{en_diss}
\end{equation}
where 
\begin{equation}
{\cal E}_{\rm R} = \frac{1}{2} m  a^{2} \dot{x}^{2} + m a^{2} n^{2} x^{2}
\end{equation}
 is the total mechanical energy of the radial oscillations and we have neglected the small terms $\eta, \zeta \ll 1$ in the l.h.s. The first term in the r.h.s. of eq.~(\ref{en_diss}), that is, $ -2bm\, a^{2} \dot{x}^{2}$, is the power dissipated by the tides inside the planet that produces a damping of the radial oscillations, while the term $m a^{2} \dot{x} f(t)$  is the power that excites the radial oscillations. The small constant force $\xi$ cannot excite any oscillation, but slightly shifts the position of equilibrium of the radial oscillator, so we shall neglect it in the following. 

Equation~(\ref{x_eq1}) is a Mathieu equation with damping and a forcing term. A first-order parametric resonance cannot occur because it would imply that the frequency of variation of the pulsation should be equal to $2 n$ which is not possible because it would imply $\Omega  = 0 $. Parametric resonances of the second or higher orders are similarly not possible because of  the dissipation term \citep{LandauLifshitz60}. Therefore, we shall consider the stationary solution of eq.~(\ref{x_eq1}) under the effect of the forcing term in the r.h.s. For the sake of simplicity, we neglect the small terms $\eta, \zeta \ll 1$ in the l.h.s. of eq.~(\ref{x_eq1}), thus assuming that the frequency of the free oscillations of $x$ is equal to the unperturbed orbital mean motion $n$. 

\subsection{Solution of the radial motion equation and power dissipated inside the planet}
\label{sol_rad_motion}

The quadrupole moment $T_{\rm s}$ is not constant because the magnetic flux tubes in the convection zone of the host star have a finite lifetime (see Sect.~\ref{tidal_def}). We quantify this effect by considering the autocorrelation function $R_{f} (\tau)$ of the forcing term on the r.h.s. of eq.~(\ref{x_eq1}), 
\begin{equation}
R_{f}( \tau ) = \langle f(t +\tau) f(t) \rangle,
\end{equation}
where $\tau$ is the time lag and the angular brackets indicate an ensemble average over many realizations of the forcing term $f(t)$. It oscillates with the frequency $\gamma \equiv 2(n-\Omega)$, that is large in comparison with the inverse of the mean lifetime of the magnetic flux tubes producing the quadrupole term $T_{\rm s}(t)$ in eq.~(\ref{fforcing}). Therefore, $R_{f}(\tau)$ can be written as the product of the autocorrelation of the quadrupole term $T_{\rm s}$ by that of the rapidly oscillating term $\cos \gamma t$. The autocorrelation of $\cos \gamma t$ is $(1/2) \cos \gamma \tau$, while that of $T_{\rm s}$ is assumed to be proportional to a decaying exponential with a typical timescale $(D/2)^{-1}$, that we take equal to the mean lifetime of the magnetic flux tubes. By combining these two contributions we obtain
\begin{equation}
R_{f}(\tau) = \langle f(t+\tau) f(t) \rangle = \frac{1}{2} A^{2}_{\rm s} \exp\left( -\frac{D}{2} | \tau | \right) \cos ( \gamma \tau).   
\label{auto_corr}
\end{equation} 
where the amplitude $A_{\rm s}$ is 
\begin{equation}
A_{\rm s} = \left(\frac{\langle T_{\rm s}^{2} \rangle}{I_{\rm s}^{2}}\right)^{1/2} A_{\rm x},
\label{fforcing_amp1}
\end{equation}
with $A_{\rm x}$ given by eq.~(\ref{fforcing_amp}). The quadratic mean relative amplitude of the stellar quadrupole $\langle T_{\rm s}^{2} \rangle / I_{\rm s}^{2}$, can be obtained from the mean filling factor $f_{\rm s}$ and intensity of the magnetic field $B_{0}$ in the overshoot layer as discussed in Sect.~\ref{model_ft}.  An estimate of the timescale $(D/2)^{-1}$ can be obtained by the rate of decrease of the amplitudes of successive peaks in the autocorrelation function of the light curve of an active star, while the lag between consecutive peaks gives a measure of the stellar rotation period \citep[cf.][]{Lanzaetal14,Gilesetal17}. 

The solution of eq.~(\ref{x_eq1}) can be obtained by considering the Fourier transforms of the functions $x(t)$ and $f(t)$. We define the Fourier transform $\hat{x}(\nu)$ of $x(t)$ as
\begin{equation}
\hat{x} (\nu) = \frac{1}{\sqrt{2\pi}} \int_{-\infty}^{\infty} x(t) \exp(- i \nu t)\, dt,  
\label{fourier_transf}
\end{equation}
where $\nu$ is the frequency and $i = \sqrt{-1}$ the imaginary unit. The inverse Fourier transform allows us to express $x(t)$ in terms of $\hat{x}(\nu)$ as
\begin{equation}
x(t) = \frac{1}{\sqrt{2\pi}} \int_{-\infty}^{\infty} \hat{x}(\nu) \exp( i\nu t) \, d\nu. 
\end{equation}
The power spectrum of $x$ can be defined as $S_{x} (\nu) = \hat{x}(\nu) \hat{x}^{*} (\nu)$, where the asterisk indicates complex conjugation. It is equal to the Fourier transform of the autocorrelation function of $x$, that is, $R_{x} (\tau) \equiv \langle x(t+\tau) x(t) \rangle $
\begin{equation}
S_{x} (\nu) = \frac{1}{\sqrt{2\pi}} \int_{-\infty}^{\infty} R_{x}(\tau) \exp(- i \nu \tau)\, d\tau.  
\label{sf_ft}
\end{equation}
The expectation value of the square of $x$, that is, $\langle x(t)^2 \rangle$, when the mean of $x$ is zero ($\langle x \rangle = 0$), can be obtained by applying the inverse Fourier transform for $\tau = 0$ because by definition $\langle x(t)^{2} \rangle = R_{x} (0)$ 
\begin{equation}
\langle x(t)^{2} \rangle = { \frac{1}{ \sqrt{2\pi}} } \int_{-\infty}^{\infty} S_{x} (\nu) \, d \nu. 
\label{x^2m}
\end{equation}
%%%%%%
By taking the Fourier transforms of both sides of eq.~(\ref{x_eq1}), neglecting the term $\xi$ and the small terms $\eta , \zeta \ll 1$ because parametric oscillations are not excited, and finally taking the complex conjugate of the transforms themselves, we find the power spectrum of $x$ as
\begin{equation}
S_{x} (\nu) \equiv \hat{x}(\nu) \hat{x}^{*} (\nu) = \frac{S_{f} (\nu)}{(n^{2}- \nu^{2})^{2} + 4 b^{2} \nu^{2}},
\label{sx}
 \end{equation}
 where $S_{f}(\nu) \equiv \hat{f}(\nu) \hat{f}^{*} (\nu)$ is the power spectrum of the forcing $f(t)$. 
 
The power spectrum of $f(t)$ is the Fourier transform of its autocorrelation function as given by eq.~(\ref{auto_corr}). 
After application of the definition (\ref{fourier_transf}) and some algebraic rearrangements, we obtain \citep[cf. the analogous system studied by][]{Zhouetal97} 
\begin{equation}
S_{f}  (\nu; \gamma, D) = \frac{A_{\rm s}^{2} D}{2\sqrt{2 \pi}} \frac{\frac{1}{4} D^{2} + \gamma^{2} + \nu^{2}}{\left( \frac{1}{4} D^{2} + \gamma^{2} - \nu^{2} \right)^{2} + D^{2} \nu^{2}},   
\label{sf}
\end{equation}
where we have explicitly indicated that  $S_{f} (\nu)$ depends also on the parameters $\gamma$ and $D$ that enter into the expression of the autocorrelation of the forcing function $f$ (cf. eq.~\ref{auto_corr}). 

The mean power $P_{\rm T}$ dissipated by the tides inside the planet is given by (cf. eq.~\ref{en_diss})
\begin{equation}
P_{\rm T} \equiv 2bm \, a^{2} \langle \dot{x}(t)^{2} \rangle, 
\label{ptm}
\end{equation}
where $\langle \dot{x}^{2}(t) \rangle$ can be obtained from its power spectrum (cf. eq.~\ref{x^2m}) 
\begin{equation}
\langle \dot{x}^{2} \rangle = { \frac{1}{ \sqrt{2\pi}} }  \int_{-\infty}^{\infty} S_{\dot{x}} (\nu) d\nu = { \frac{1}{ \sqrt{2\pi}} }  \int_{-\infty}^{\infty} \nu^{2} S_{x} (\nu) \, d\nu, 
\end{equation}
because the Fourier transform of the time derivative is $\hat{\dot{x}} = i\nu \hat{x}$ that implies $S_{\dot{x}} (\nu) = \nu^{2} S_{x} (\nu)$. 
By substituting eq.~(\ref{sx}), we obtain
\begin{eqnarray}
P_{\rm T} & = & 2b m a^{2}  \frac{1}{ \sqrt{2\pi}}  \int_{-\infty}^{\infty}  \frac{ \nu^{2} S_{f} (\nu; \gamma, D)}{(n^{2}- \nu^{2})^{2} + 4 b^{2} \nu^{2}} \, d\nu \label{pt1} \\
&  \simeq & 2b m a^{2}  \frac{1}{ \sqrt{2\pi}} S_{f} (n; \gamma, D) \int_{-\infty}^{\infty}  \frac{ \nu^{2} d\nu}{(n^{2}-\nu^{2})^{2} + 4 b^{2} \nu^{2}} \label{pt2} \\
& = & 2b m a^{2}  \frac{1}{ \sqrt{2\pi}} \frac{\pi}{2b} S_{f} (n; \gamma, D) = \sqrt{\frac{\pi}{2}} m  a^{2} S_{f} (n; \gamma, D) \label{pt3},
\end{eqnarray}
where we made use of the fact that the denominator in the integrand of eq.~(\ref{pt1}) becomes very small when $\nu$ is very close to $n$ because $b \ll n$. Therefore, we can take the factor $S_{f} (\nu; \gamma, D)$ out of the integral by computing it for $\nu = n$ and integrate the remaining factor by applying the theorem of the residues (see Appendix~\ref{math_app}) to obtain eq.~(\ref{pt3}). Equation~(\ref{pt3}) shows that the mean dissipated power $P_{\rm T}$ is independent of the dissipation rate $b$, which is a very useful result in view of our ignorance of the internal structure and rheology of the planet determining its modified tidal quality factor (cf. Sect.~\ref{planet_tidal_diss}).  

The value of $S_{f}(n; \gamma, D)$ is given by eq.~(\ref{sf}) with $\nu =n$. It reaches its maximum  when $\gamma \equiv \gamma_{\rm res}= \sqrt{n^{2} - D^{2}/4}$, that is, $2(n-\Omega) \simeq n$ or $\Omega \simeq n/2$ because $D \ll n$ in very active stars in which active longitudes persist for timescales ranging from hundreds of days up to a decade and the orbital period of the planet is less than a few days.  Therefore, the maximum of the dissipated power at the resonance is sharp and much greater than the power dissipated far from the resonance. By combining eqs.~(\ref{sf}) and~(\ref{pt3}), the maximum dissipated power is
\begin{equation}
P_{\rm T \max} = \frac{1}{2} \, \frac{m a^{2} A_{\rm s}^{2}}{D},
\end{equation}
where we made use of the fact that $D \ll n$. The power decreases rapidly from its maximum when $\gamma$ varies around the resonance value $\gamma_{\rm res}$. A power of one half of  $P_{\rm T \max}$ is obtained for $\gamma \sim \gamma_{\rm res} [1 \pm D/(2n)]$, considering eq.~(\ref{sf}) and  that $D \ll n$. 

The angular velocity  $\Omega$ will change during the evolution of the star in its pre-main-sequence and main-sequence phases owing to the changes in the stellar moment of inertia and the loss of angular momentum carried away by the stellar magnetized wind. Therefore, the value of $P_{\rm T}$ will vary as $\Omega$ evolves and crosses the resonance at $\Omega \simeq n/2$. As we shall see in Sect.~\ref{applications}, the crossing of the resonance occurs over a time interval much shorter than the internal timescale of heat redistribution inside the planet, so that all the power dissipated during the resonance is stored into the planet and released on a much longer timescale. Since the resonance is sharp because $D \ll n$ (see above), the total energy $E_{\rm D}$ dissipated inside the planet during the crossing of the resonance can be evaluated as
\begin{equation}
E_{\rm D} \equiv \int_{-\infty}^{\infty} P_{\rm T} (t) \, dt \simeq \sqrt{\frac{\pi}{2}} m a^{2} \left(\frac{dt}{d\gamma}\right)_{\rm res} \int_{-\infty}^{\infty} S_{f} (n; \gamma, D) \, d\gamma, 
\label{ed}
\end{equation}
 where $dt/d\gamma$ is the inverse of the rate of change of the frequency of the forcing term $f$ that is produced by the evolution of the stellar rotation and of the orbit semimajor axis under the action of the tides inside the star (cf. Sect.~\ref{braking_and_tides}). Such an inverse rate of change is evaluated at the resonance, that is, for $\gamma = \sqrt{n^{2} - D^{2}/4}$,  and it is taken out of the integral thanks to the sharpness of the resonance. The integral can be evaluated with the theorem of the residues (see Appendix~\ref{math_app}, eq.~\ref{energy_integral}) giving 
\begin{equation}
E_{\rm D} =   \frac{\pi}{2} \, m a^{2} A_{\rm s}^{2} \left( \frac{dt}{d\gamma}\right)_{\rm res}.  
\label{ediss_rad}
\end{equation} 
It is remarkable that $E_{\rm D}$ is independent of $D$ and is a function only of $A_{\rm s}$, that is, of the non-axisymmetric component of the stellar quadrupole moment $T_{\rm s}$, the orbital parameters, and the inverse rate of change of $\gamma$ at the resonance, that is, of the evolution of the angular momentum of the system. The amount of dissipated energy decreases rapidly with the increase of the orbit semimajor axis, scaling as $E_{\rm D} \propto a^{-8}$ (cf. eqs.~\ref{fforcing_amp}, \ref{fforcing_amp1}, and~\ref{ediss_rad}), which implies that this heating mechanism is mostly limited to operate in planets with ultra-short-period orbits ($P_{\rm orb} \la 1$~days) around young active late-type stars. 

%\subsubsection{Variation of the orbital elements}
%The radial motion that is excited by the stellar non-axisymmetric quadrupole moment when the angular velocity of rotation of the star $\Omega \simeq n/2$, has virtually the same period of the orbital motion because the terms $\eta, \zeta, \xi \ll 1$ in equation~(\ref{x_eq}). The corresponding variations in the orbital elements, specifically in the orbital semimajor axis $a$ and the eccentricity $e$, average to zero along one complete orbit. 

%This can be shown by computing the variations in the total orbital energy $E_{\rm orb} = G m_{\rm s} m_{\rm p} /(2a)$ and in the specific angular momentum $h = \sqrt{Gm_{\rm T} a (1-e^{2})}$ due to the perturbing gravitational quadrupole potential of the star and integrating them along the unperturbed orbit \citep[e.g.][]{MardlingLin02}. We do not provide the details of the calculations here, but consider the main consequence for the tidal heating. Specifically, the fact that there is no variation either in the eccentricity or in the orbital semimajor axis along a complete orbit implies  that there is no additional internal heating of the planet due to a variation of its orbital elements in addition to that discussed in the previous section. 

\subsection{Libration of the planet}
\label{planet_libration}
Now we focus on the equations~(\ref{f_eq}), (\ref{phi_eq}), and~(\ref{psi_eq}) for the angular variables and consider, for simplicity, the case of a circular orbit. As we shall see, such an assumption is justified because the libration of the planet that we want to investigate occurs far from the resonance that excites  the radial motion. In such a regime, $x(t)$ is damped  on a timescale much shorter than the lifetime of the system (see Sect.~\ref{applications}) justifying the assumption. For a general discussion of the effects of librations in the case of telluric planets, see, for example, \citet{LaBars16} and references therein. 

Indicating with $a$ the constant orbital radius $r$, eqs.~(\ref{f_eq}), (\ref{phi_eq}), and~(\ref{psi_eq}) read
\begin{eqnarray}
\ddot{f} +\frac{3}{2} n^{2} \left[ \left( \frac{T_{\rm s}}{m_{\rm s} a^{2}} \right) \sin 2 \alpha + \left( \frac{T_{\rm p}}{m_{\rm p} a^{2}} \right) \sin 2 \beta \right] = 0, \label{fdd_eq}\\
\ddot{\varphi} - \frac{3}{2} n^{2} \left( \frac{m_{\rm p}}{m_{\rm T}} \right) \left( \frac{T_{\rm s}}{I_{\rm s}} \right) \sin 2 \alpha = 0, \label{phidd_eq} \\
\ddot{\psi} - \frac{3}{2} n^{2} \left( \frac{m_{\rm s}}{m_{\rm T}} \right) \left( \frac{T_{\rm p}}{I_{\rm p}} \right) \sin 2 \beta = 0, \label{psidd_eq} 
\end{eqnarray}
where we made use of Kepler III law ($n^{2} a^{3} = Gm_{\rm T})$ and introduced the angle $\alpha \equiv f- \varphi = (n -\Omega) t$ and the libration angle of the planet $\beta = f -\psi$. By subtracting eq.~(\ref{psidd_eq}) from eq.~(\ref{fdd_eq}), we obtain an equation for $\beta$ as
\begin{equation}
\ddot{\beta} + \frac{1}{2} \omega_{\rm p}^{2} \sin 2 \beta = -\frac{3}{2} n^{2} \left( \frac{T_{\rm s}}{m_{\rm s}a^{2}} \right) \sin \gamma t, 
\label{beta_eq}
\end{equation}
where $\gamma = 2 (n-\Omega)$ is the frequency of the forcing introduced in Sect.~\ref{sol_rad_motion}. This is the equation of a forced simple pendulum with pulsation $\omega_{\rm p}$ given by
\begin{equation}
\omega_{\rm p}^{2} = 3 n^{2} \left[ \frac{1}{m_{\rm p} a^{2}} + \left( \frac{m_{\rm s}}{m_{\rm T}} \right) \frac{1}{I_{\rm p}} \right] T_{\rm p} \simeq 3 n^{2} \left( \frac{T_{\rm p}}{I_{\rm p}} \right),
\label{omegap}
\end{equation}
because the moment of inertia of the planet $I_{\rm p}$ is much smaller than $m_{\rm p} a^{2}$ and $m_{\rm s} \simeq m_{\rm T}$. 

The effect of tides is not included into eq.~(\ref{beta_eq}) as it tends to synchronize the rotation of the planet with the orbital motion. In other words, tides produce a damping of $\dot{\beta}$ over a timescale $\tau_{\rm rot}$ given by \citep[cf. eq.~9 in][]{Guetal03}
\begin{equation}
\tau_{\rm rot}^{-1} = \left| \frac{\ddot{\psi}}{n-\dot{\psi}} \right| = \left| \frac{\ddot{\beta}}{\dot{\beta}} \right| = \frac{9n}{2h_{\rm p} Q^{\prime \rm T}_{\rm p}} \left( \frac{m_{\rm s}}{m_{\rm p}} \right) \left( \frac{R_{\rm p}}{a} \right)^{3},
\label{libr_damping}
\end{equation}
where we considered that $\ddot{\beta} = - \ddot{\psi}$ because the orbital motion is not appreciably perturbed given that $I_{\rm p} \ll ma^{2}$ and defined $h_{\rm p} \equiv I_{\rm p} /(m_{\rm p} R_{\rm p}^{2})$, while the other symbols have already been introduced above. For the Earth, $h_{\rm p} = 0.3307$. 

The equation for the libration angle $\beta$, including  tidal effects  and in the limit of a small amplitude of the libration, that is, $\beta \ll 1$, becomes
\begin{equation}
\ddot{\beta} + 2b_{\ell} \,\dot{\beta} + \omega_{\rm p}^{2} \, \beta = -\frac{3}{2} n^{2} \left( \frac{T_{\rm s}}{m_{\rm s}a^{2}} \right) \sin \gamma t, 
\label{beta_eq_com}
\end{equation}
where $ 2 b_{\ell} \equiv \tau_{\rm rot}^{-1}$ is the damping constant. 

Equation~(\ref{beta_eq_com}) shows that the libration of the planet can be resonantly excited when  $\gamma = 2(n-\Omega) $ is close to $\omega_{\rm p}$ with the width of the resonance controlled by the tidal dissipation through the parameter $b_{\ell}$. Considering eq.~(\ref{omegap}) and that $T_{\rm p} \sim (10^{-3} -10^{-4}) I_{\rm p}$ (cf. eq.~\ref{remus_teq}), we have that $\omega_{\rm p} \ll n$, thus the resonance occurs only when $\Omega \sim n$, that is, the angular velocity of rotation of the star is close to the mean orbital motion of the planet. On the other hand, we saw in Sect.~\ref{radial_motion} that the radial motion is excited when $\Omega \sim n/2$. Therefore, the libration resonance is well separated from the resonance that excites the radial motion provided that the damping parameters $b$ and $b_{\ell}$ are much smaller than $n$ and $\omega_{\rm p}$, respectively. 

\subsection{Power dissipated by librations inside the planet}
\label{libration_power}

The power $P_{\rm libr}$ dissipated by the tides inside a librating planet can be written as \citep[][Sect.~2.2]{Ogilvie14}
\begin{equation}
P_{\rm libr} = \hat{\omega} T_{\rm t},
\end{equation}
where $\hat{\omega} = 2 (\dot{f} - \dot{\psi}) = 2 \dot{\beta}$ is the frequency of the semidiurnal tide acting on the planet and $T_{\rm t}$ is the tidal torque acting on the planet spin that we write as $T_{\rm t} = - 2b_{\ell} I_{\rm p} \dot{\beta}$. Therefore, the mean dissipated power becomes
\begin{equation}
P_{\rm libr} = 4 b_{\ell} I_{\rm p} \langle \dot{\beta}^{2} \rangle.
\end{equation}
Following the same procedure illustrated in Sect.~\ref{sol_rad_motion}, we compute the power spectrum of $\dot{\beta}$ from the power spectrum of the forcing term and assume that $b_{\ell} \ll \omega_{\rm p}$ to find
\begin{equation}
P_{\rm libr} = \sqrt{2 \pi} \, I_{\rm p} S_{f_{\beta}} (\omega_{\rm p}; \gamma, D), 
\end{equation}
where the power spectrum of the forcing term 
\begin{equation}
f_{\beta} \equiv -\frac{3}{2} n^{2} \left[ \frac{T_{\rm s}(t)}{m_{\rm s}a^{2}} \right] \sin \gamma t , 
\label{fbeta}
\end{equation}
is similar to that of $f$ in eq.~(\ref{sf}), the only difference between $f$ and $f_{\beta}$ being the factor $(3\Omega-n)/\gamma$ appearing in the amplitude of $f(t)$ that changes the amplitude of the power spectrum accordingly (cf. eqs.~\ref{fforcing} and~\ref{fforcing_amp}). In eq.~(\ref{fbeta}), we made the dependence of $T_{\rm s}$ on the time explicit to allow an easier comparison with eq.~(\ref{fforcing}).

The total energy $E_{\beta}$ dissipated in the planet when the frequency $\gamma$ passes across the resonance that excites the librations is obtained by integrating the power $P_{\rm libr}$ over the time. Following the same approach as in Sect.~\ref{sol_rad_motion}, we find
\begin{equation}
E_{\beta} = \pi \, I_{\rm p} \, A_{\beta}^{2} \left( \frac{dt}{d\gamma} \right)_{\beta \, \rm res},
\label{ediss_libr}
\end{equation}
where $A_{\beta}$ is the amplitude of the power spectrum in eq.~(\ref{sf}) in the case of the forcing (\ref{fbeta}) and the derivative $dt/d\gamma$ is computed at the resonance, that is, for $\gamma = \gamma_{\beta \, \rm res} \equiv \sqrt{\omega_{\rm p}^{2} - D^{2}/4}$. 

By comparing eqs.~(\ref{ediss_rad}) and~(\ref{ediss_libr}) and noting that the amplitudes $A_{\rm s}$ and $A_{\beta}$ as well as the derivatives $(dt/d\gamma)$ at the respective resonances are comparable with each other, we find that the ratio of the total dissipated energies is $E_{\beta} / E_{\rm D} \sim 2 I_{\rm p}/(ma^{2})  \sim 10^{-5} - 10^{-4} \ll 1$. Therefore, we shall neglect the energy dissipated by the librations inside the planet in comparison with the energy dissipated by the radial oscillations discussed in Sect.~\ref{sol_rad_motion}.

\subsection{Long-term evolution of the stellar rotation and orbital semimajor axis}
\label{braking_and_tides}

The quadrupole moments of the star and the planet produce a spin-orbit coupling that  exchanges angular momentum between the orbital motion and the rotation of the bodies. In the most extreme cases, this effect produces a modulation of the orbital period with a relative amplitude $\Delta P_{\rm orb}/P_{\rm orb} \la 10^{-6}$. These angular momentum exchanges are cyclic, therefore, they neither affect the long-term evolution of the orbit nor of the stellar rotation. More details can be found in \citet{Lanza20a,Lanza20b}, but they do not contribute a significant tidal dissipation inside a solid planet, so we shall neglect them.

The long-term evolution of the system is driven by the loss of angular momentum in the stellar magnetized wind and by the star-planet tidal interactions.  Here the angular momentum loss is parametrized following the simple model by \citet{Bouvieretal97} and assuming that the star is rotating as a rigid body after settling on the zero-age main sequence (ZAMS). The internal structure of the star is assumed to be steady, thus its moment of inertia $I_{\rm s}$ is constant during the evolution of its rotation. Such  approximations are useful for our illustrative applications in Sect.~\ref{applications} because we shall focus on the first $\sim 100$~Myr of evolution of the stellar rotation after the ZAMS when they are verified rather well. More sophisticated models can be applied for a detailed comparison with the observations \citep[e.g.,][]{Amardetal16,Galletetal18,Benbakouraetal19}, but this is postponed to later works because here we focus on the physical mechanism producing the internal heating of the planet.  

The angular momentum loss rate $dL_{\rm w}/dt$ due to the magnetized wind is described by 
\begin{equation}
\frac{dL_{\rm w}}{dt} = \left\{ \begin{array}{r}
                            -K_{\rm 	w} \Omega^{3} \left( \frac{R_{\rm s}}{R_{\odot}} \right)^{1/2} \left( \frac{m_{\rm s}}{M_{\odot}} \right)^{-1/2}  \; \; \mbox{for $\Omega < \Omega_{\rm sat} (m_{\rm s}) $,} \\
                            -K_{\rm w} \Omega_{\rm sat}^{2} \Omega \left( \frac{R_{\rm s}}{R_{\odot}} \right)^{1/2} \left( \frac{m_{\rm s}}{M_{\odot}} \right)^{-1/2}  \; \; \mbox{for $\Omega \geq  \Omega_{\rm sat} (m_{\rm s}) $, }
                         \end{array} \right. 
\label{dLwdt}
\end{equation}
where $K_{\rm w} = 2.7 \times 10^{40}$ kg m$^{2}$ s$^{-1}$, and $\Omega_{\rm sat}$ is the angular velocity corresponding to the saturation of the wind angular momentum loss rate that is a function of the star mass $m_{\rm s}$. We shall assume $\Omega_{\rm sat} = 14, 8, 3\; \Omega_{\odot}$ for a stellar mass $m_{\rm s} = 1.0, 0.8, 0.5\; M_{\odot}$, respectively, and linearly interpolate for masses in between.  

The tidal torque acting on the stellar rotation $\Gamma_{\rm t}$  can be expressed in terms of the parameters of the system following, e.g.,  \citet{MardlingLin02}. After some re-arrangement of the terms and taking into account also the torque produced by the stellar wind, we can write the equation for the evolution of the stellar rotation as
\begin{equation}
I_{\rm s} \frac{d\Omega}{dt} = \Gamma_{t} + \frac{dL_{\rm w}}{dt} = 9 \frac{Gm_{\rm p}^{2}}{a} \left( \frac{R_{\rm s}}{a} \right)^{5} \frac{1}{Q^{\prime \rm T}_{\rm s}} \left( \frac{n-\Omega}{n} \right) +
\frac{dL_{\rm w}}{dt}, 
\label{domegadt}
\end{equation}
where all the symbols have already been defined. 

The stellar modified tidal quality factor $Q^{\prime \rm T}_{\rm s}$ is remarkably uncertain {because it depends on the processes that dissipate the kinetic energy of the tides raised inside the star by the orbiting planet. 

Traditionally, the tidal response of a star is decomposed into a large-scale nearly-hydrostatic deformation, called the equilibrium tide, and the various kinds of waves that are excited by the time-dependent tidal potential in a reference frame rotating with the star, the so-called {dynamical tide}. In the case of late-type stars with close-by planets, the {dynamical tide} consists of the internal gravity waves excited inside their radiative zones and the inertial waves excited into their convection zones when the tidal frequency $ |\hat{\omega} |\equiv |2(n-\Omega) |  \leq 2\Omega$ \citep[see ][]{Barker20}. 
The excitation and dissipation of {dynamical tides} are associated with a remarkably larger tidal dissipation than in the case of the equilibrium tide and correspond to a decrease of the modified tidal quality factor $Q^{\prime \rm T}_{\rm s}$ by $2-4$ orders of magnitude. 

Internal gravity waves become relevant when their amplitude increases beyond the threshold required for their efficient dissipation inside the radiative zone of the star, that requires a planet mass remarkably greater than the mass of Jupiter in the case of late-type stars of mass $0.7-1.1$ solar masses, ages younger than a few Gyr, and orbital periods of the order of 1~day or shorter as in our case; see, e.g.,  Section~6 of \citet{Barker20} {or \citet{Ahuiretal21}}. Therefore, their dissipation does not appreciably contribute to the evolution of the systems we consider here. 

On the other hand, inertial waves in the convection zone, the restoring force of which is the Coriolis force, play a relevant role in our case in the range of tidal frequencies where they are excited. The associated $Q^{\prime \rm T}_{\rm s}$ shows a highly erratic dependence on the tidal frequency for given stellar parameters, therefore, it is useful to adopt an average value to compute the evolution of star-planet systems as in, e.g., \citet{Mathis15} or \citet{Galletetal17}. Such an average value is generally $2-3$  orders of magnitude smaller than that associated with the equilibrium tide and is proportional to the inverse of the square of the stellar angular velocity because the Coriolis acceleration is proportional to $\Omega$, viz., $Q^{\prime \rm T}_{\rm s} \propto \Omega^{-2}$ \citep{OgilvieLin07, Ogilvie13, Barker20}. }

In view of these results of the tidal theory, we 
assume
 \begin{equation}
 Q^{\prime \rm T}_{\rm s} = Q_{\rm s 0}^{\prime \rm T} (\Omega_{0}/ \Omega_{\rm ref})^{-2}
 \label{qpts}
 \end{equation}
in the regime of tidal frequency where inertial waves are excited with $Q^{\prime \rm T}_{\rm s0} = 4 \times 10^{7}$ for $m_{\rm s} \geq 0.9$~M$_{\odot}$ or $Q^{\prime \rm T}_{\rm s0} = 2 \times  10^{7}$ for $m_{\rm s} < 0.9$~M$_{\odot}$, $\Omega_{0}$ being the angular velocity of the star on its arrival on the ZAMS, and $\Omega_{\rm ref}$ the angular velocity when the modified tidal quality factor is equal to $Q^{\prime \rm T}_{\rm s0}$; we assume it to correspond to a rotation period of 10~days, viz. $\Omega_{\rm ref} = 7.27 \times 10^{-6}$~s$^{-1}$. {The values of $Q^{\prime \rm T}_{\rm s}$ given by eq.~(\ref{qpts}) are similar within a factor of $\approx 1-3$ to those computed by \citet{Mathis15} and \citet{Galletetal17} with an approximate two-layer model for late-type stars on the main sequence, while they are about a factor of $2-5$ greater than the values computed by \citet{Barker20} with {stratified models}. In view of the significant uncertainties in the present tidal theory, a closer agreement between their estimates and ours is not pursued here.   }

When inertial waves are not excited, we increase the modified tidal quality factor  $Q^{\prime T}_{\rm s}$ by a factor of the order of $10^{3}$, thus producing a much smaller tidal coupling between the stellar spin and the orbit \citep[cf.][]{OgilvieLin07,Barker20}. More precisely, {given the present uncertainties of the equilibrium tide theory to be applied in that regime \citep[see, e.g., Section~2 of][]{Barker20}, we take the freedom of adjusting} the precise value of $Q_{\rm s}^{\prime \rm T}$ to reproduce the current orbital period and rotation period of the star in the modelled systems starting from their adopted initial conditions (see Sect.~\ref{applications} for details). 

The evolution of the orbit semimajor axis $a$ can be computed from the tidal torque $\Gamma_{\rm t}$ considering the conservation of the total angular momentum of the system. After some algebra, we find
\begin{equation}
\frac{da}{dt} = \frac{18}{Q^{\prime \rm T}_{\rm s}} \left( \frac{R_{\rm s}}{a} \right)^{5} \left( \frac{m_{\rm p}}{m_{\rm s}} \right) \left( \frac{\Omega - n }{n} \right) n a.  
\label{dadt}
\end{equation}
In the next section, we shall make use of eqs.~(\ref{domegadt}) and~(\ref{dadt}) to compute some illustrative models of the evolution of the stellar rotation and the semimajor axis for  some planetary systems in order to show how the conditions for the excitation of the radial oscillations of the planet can be realized along their evolution. 

We shall focus on the dissipation inside the planet considering the evolution from the ZAMS. The angular velocity of the star on the ZAMS depends on its past evolution, in particular on its contraction during the pre-main-sequence phase, and the particular moment when the mechanical coupling with its circumstellar disc came to an end leaving the contracting protostar free to spin up. Therefore, the initial rotation period $2\pi/\Omega_{0}$ on the ZAMS is strongly dependent on the protostar mass and the time when the decoupling from the disc occurred and can range from a few hours up to several days in the case of a long-lasting disc locking \citep[e.g.,][]{Bouvieretal97,GalletBouvier15}. In view of our ignorance on the pre-main-sequence evolution of our stars, we shall regard $\Omega_{0}$ as a free parameter and adjust it to obtain the resonant excitation of the radial motion at some time after the ZAMS. 

\section{Applications}
\label{applications}

We consider three MESA stellar models \citep[see][and references therein]{Fieldsetal18,Paxtonetal19} to evaluate the non-axisymmetric quadrupole moment for three late-type active stars with masses typical of those hosting ultra-short-period planets. The  non-axisymmetric quadrupole moment $T_{\rm s}$  is computed according to eq.~(\ref{qandt}) considering $\theta_{0} = 30^{\circ}$ that corresponds to a filling factor $f_{\rm s} \simeq 0.07$ of the whole stellar surface. The corresponding relative amplitude of the optical light modulation of the star is $\approx 25$ percent that is typical of late-type stars with a rotation period close to $\approx 1$~day and younger than $100$~Myr as we shall consider in our applications \citep[e.g.,][]{Messina21}. 

The integral $\cal J$ in eq.~(\ref{j_integ}) is evaluated by considering the internal structure model of the host star as computed by means of the MESA Web interface\footnote{http://mesa-web.asu.edu/}. We assume a metallicity $Z=0.02$, a ratio of the mixing length to the pressure scale height equal to $2.0$, and standard values for the other parameters and options. We list the stellar parameters in the cases of models of mass $0.7, 0.8$ and $0.9$~M$_{\odot}$ in Table~\ref{star_quadrupole_table} at the indicated ages that correspond to models during the main-sequence phase of their evolution. Given their slow evolution along the main sequence, these models can be regarded as representative of the stellar internal structures all along this phase for our purposes. 

The magnetic field $B_{0}$ in the overshoot layer immediately below the base of the convection zone at $r=r_{\rm b}$ is chosen to have $T/I_{\rm s} = 10^{-6}$, where $I_{\rm s}$ is the stellar moment of inertia. Those values of $B_{0}$ are comparable with the magnetic field instability thresholds for the emergence of the flux tubes stored in the overshoot layers of late-type, rapidly rotating stars according to the model of \citet{Granzeretal00}. The upper limit $r_{\rm L}$ to which the integration in eq.~(\ref{j_integ}) is extended is also listed in Table~\ref{star_quadrupole_table} as well as the magnetic energy $E_{\rm mag}$ of the vertical flux tube for $f_{\rm s} = 0.07$ according to the model of \citet{Lanza20a}. 
%%%%%%%%%%%%%%%%%%%%%%%%%%
\begin{table}
\caption{Stellar models considered to compute the non-axisymmetric quadrupole moment $T_{\rm s}$. The intensity $B_{0}$ of the magnetic field of the flux tube of angular radius $\theta_{0} = 30^{\circ}$ in the overshoot layer at $r=r_{\rm b}$  is chosen to have $T_{\rm s}/I_{\rm s} = 10^{-6}$, while the magnetic energy of the flux tube is $E_{\rm mag}$ (see text). }
\begin{center}
\begin{tabular}{cccc}
\hline 
Star mass (M$_{\odot}$) & 0.7 & 0.8 & 0.9 \\
Star radius (R$_{\odot}$) & 0.653 & 0.721 & 0.888 \\
Age (Myr) & 583 & 256 & 7452 \\
$r_{\rm b}/R_{\rm s}$ & 0.676 & 0.690 & 0.694 \\
$r_{\rm L}/R_{\rm s}$ & 0.990 & 0.990 & 0.987 \\
$\cal J$ ($10^{-4}$ s$^{2}$~m$^{-1}$) & 5.9784 & 6.0804 & 7.9771 \\ 
$B_{0}$ (T) & 71.0 & 66.0 & 50.0 \\ 
$E_{\rm mag}$ ($10^{34}$~J) & 1.56 & 1.83 & 1.96 \\ 
\hline 
\end{tabular}
\end{center}
\label{star_quadrupole_table}
\end{table}
%%%%%%%%%%%%%%%%%%%%%%%%%%%%%%%%%%%%%

Three illustrative applications to ultra-short-period planets are introduced in this Section, specifically  to \object{CoRoT-7b}, \object{Kepler-78b}, and \object{K2-141b}, respectively. They are not intended as detailed and accurate models of the evolution of these systems, but only as examples to show the typical powers, total energies, and timescales involved in the heating mechanism introduced in this paper. Therefore, we shall not compare the results of our models with the observations in detail. In the case of CoRoT-7 and K2-141, another planet has been detected in the system with an orbital period longer than that of the considered innermost planet. 

The presence of outer companions may be a common property of ultra-short-period planets and could have played a decisive role in their migration close to the host star \citep[cf.][]{Daietal18,Petrovichetal19}. For simplicity, we shall not consider the effects due to such companions and model the evolution assuming that these systems consist only of the host star and their closest planet. 

The parameters assumed for the considered systems are listed in Table~\ref{systems_table}. To model their evolution starting from the ZAMS corresponding to the time $t=0$, we adopt initial values of the orbit semimajor axis $a(t=0)$, orbital period $P_{\rm orb} (t=0)$, and stellar rotation period $P_{\rm rot} (t=0) = 2\pi/\Omega_{0}$ as listed in Table~\ref{systems_models}. We start from rapidly rotating stars on the ZAMS and do not consider the evolution of the radius of the host star along its main-sequence lifetime.  On the other hand, had the star settled on the main sequence as a slow rotator with an initial  $P_{\rm rot} > 2 P_{\rm orb}$, our mechanism could not operate because the spin-orbit resonance responsible for the excitation of the radial oscillations of the planetary orbit and the consequent dissipation could not occur. %Therefore, the assumption that the host star started its evolution as a fast rotator on the ZAMS with an initial $P_{\rm rot}$ comparable with the short orbital period of the planet $P_{\rm orb}$ is required for the operation of our mechanism. 

Another assumption of our model is that the planet was on a very close orbit since the star started its main-sequence evolution as could happen in the case of a migration occurred inside the protoplanetary disc or an early inward scattering in multiplanet systems. Other mechanisms that could bring the planet very close to its host star after a much longer time should not produce a heating of the planet by our mechanism because the rotation of the star would have already been braked below the value required to excite the radial motion resonance \citep[e.g.][]{LanzaShkolnik14}. 

The angular velocity $\Omega_{\rm sat}$ corresponding  to the saturation of the angular momentum loss in the stellar wind is computed by a linear interpolation considering its dependence on the mass of the host star (cf. Sect.~\ref{braking_and_tides}). We consider a fixed amplitude of the non-axisymmetric quadrupole moment $T_{\rm s}/I_{\rm s} =10^{-6}$ because it is relevant only close to the resonance that excites the radial orbital oscillations of the planet. The lifetime $(D/2)^{-1}$ of the magnetic flux tubes responsible for the non-axisymmetric quadrupole is assumed to be $500$~days in all the cases (cf. Sect.~\ref{tidal_def}). 

\begin{figure}
%\hspace*{-7mm}
 \centering{
 \includegraphics[width=8cm,height=10cm,angle=90]{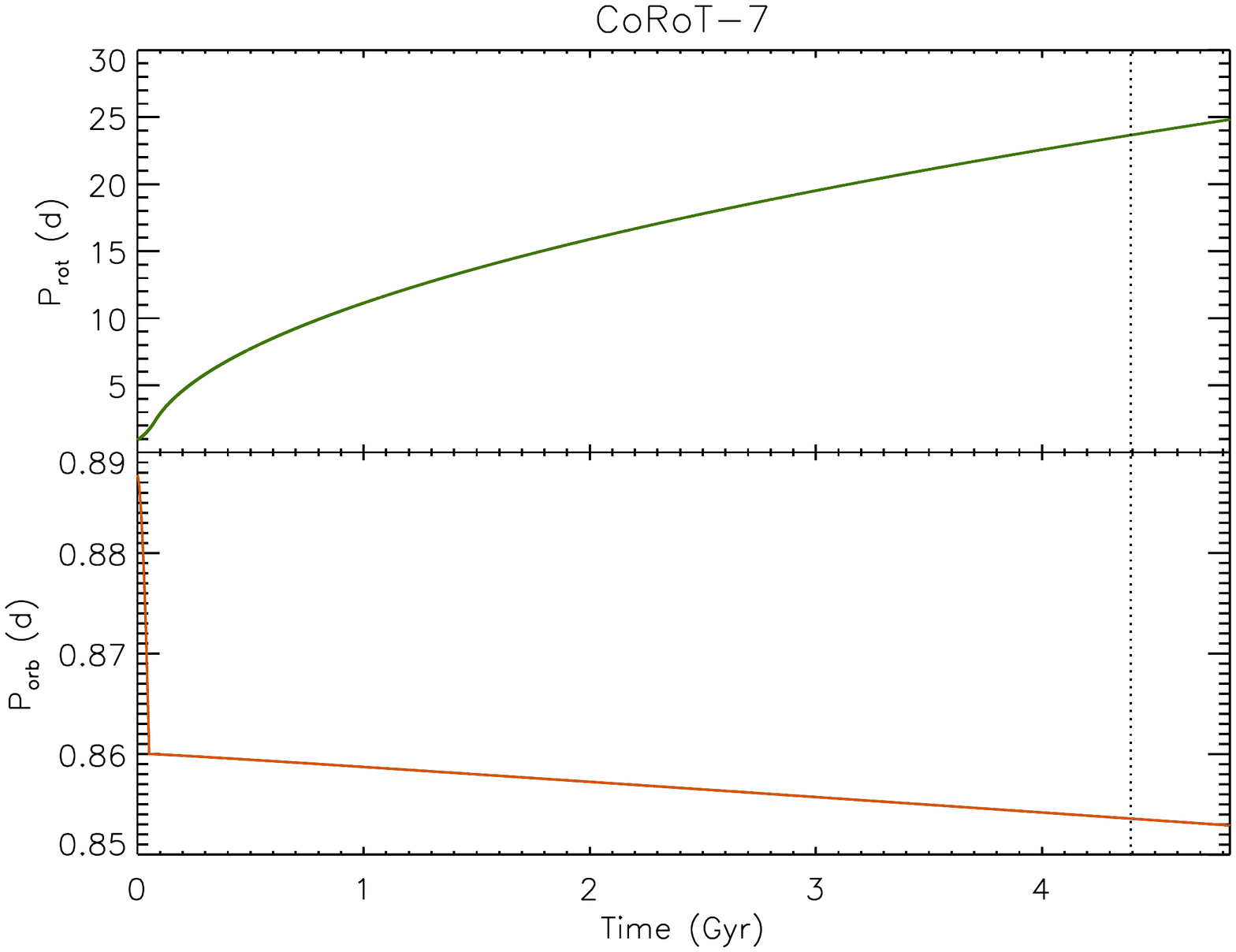}} %CoRoT-7_evol_corr.pdf}} % HD106252_1_rhkcorr_KR_final.pdf}}
% \vspace*{-10mm}
   \caption{Upper panel: The  rotation period of CoRoT-7 vs. the time as obtained from our model of the stellar angular momentum and tidal evolution introduced in Sect.~\ref{braking_and_tides}. The vertical dotted line marks the time corresponding to the presently measured mean rotation period of the star that should correspond to its age in the hypothesis that its rotational evolution followed our model. Lower panel: The orbital period of CoRoT-7b vs. the time as computed with our tidal model in Sect.~\ref{braking_and_tides}. The vertical dotted line marks the current value of the orbital period.}
              \label{corot7_evol}%
\end{figure}

The stellar modified tidal quality factors are listed in Table~\ref{systems_models} for the case when inertial waves are excited (IW) and for the case when only the equilibrium tide (ET) produces dissipation inside the star as discussed in Sect.~\ref{braking_and_tides}. The value of $Q^{\prime \rm T}_{\rm s}$ in the case of the equilibrium tide is adjusted to reproduce both the observed current orbital period and stellar rotation period as indicated by the vertical dotted lines in Figs.~\ref{corot7_evol}, \ref{kepler78_evol}, and~\ref{k2-141_evol}.  The system ages estimated with our model at those epochs are listed in Table~\ref{systems_models}. They can differ from those in the literature as reported in Table~\ref{systems_table} because of the large uncertainty (up to $30-40$ percent) in those estimates, mainly based on gyrochronology, as well as the simplifying assumptions adopted in our illustrative models. 

\begin{figure}
%\hspace*{-7mm}
 \centering{
 \includegraphics[width=8cm,height=10cm,angle=90]{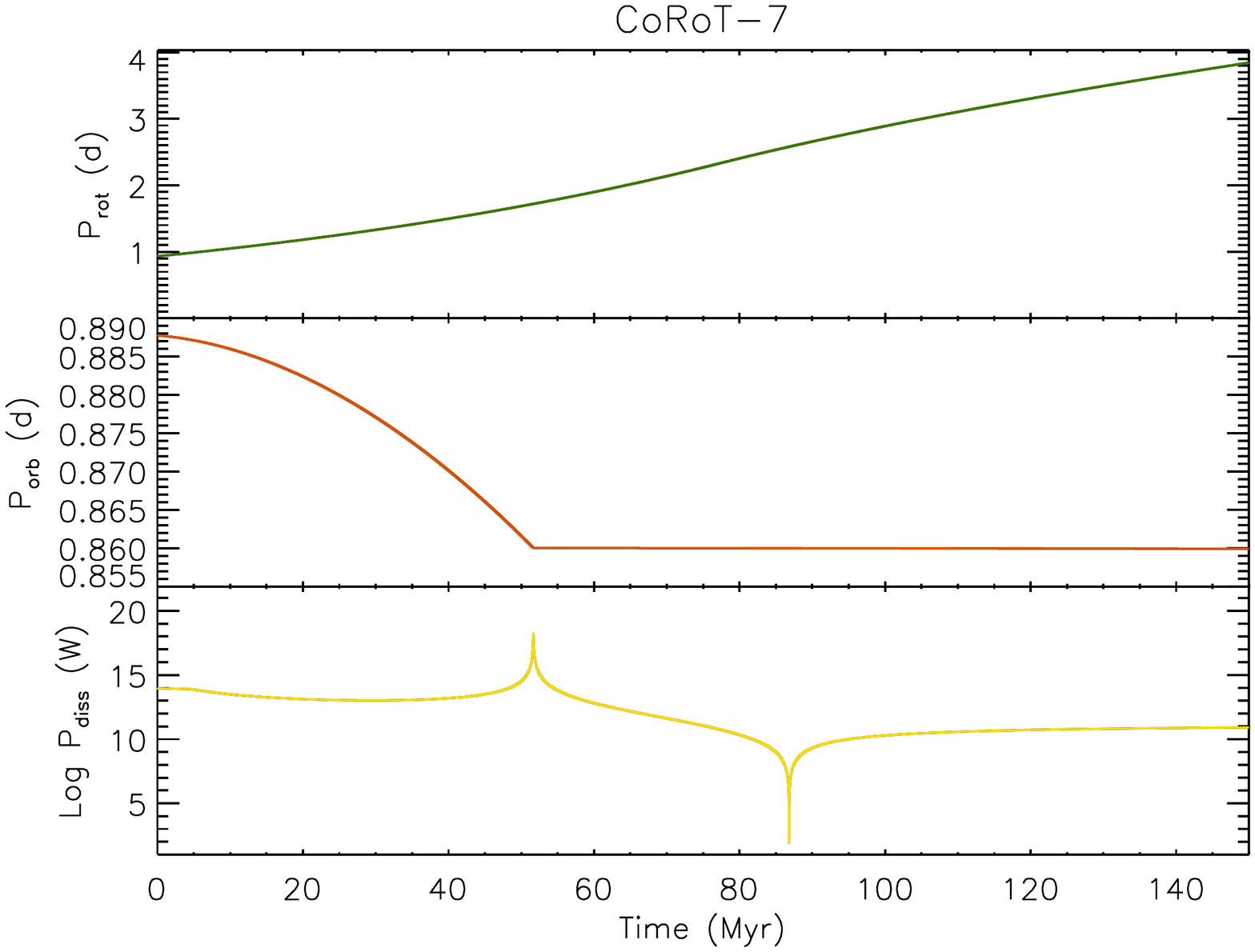}} % CoRoT-7_detail_corr.pdf}} % HD106252_1_rhkcorr_KR_final.pdf}}
% \vspace*{-10mm}
   \caption{Upper panel: An enlargement of the upper plot of Fig.~\ref{corot7_evol} showing the rotation period of CoRoT-7 vs. the time as obtained from our model of the stellar angular momentum and tidal evolution introduced in Sect.~\ref{braking_and_tides}. Middle panel: an enlargement of the lower plot of Fig.~\ref{corot7_evol}, showing the orbital period of CoRoT-7b vs. the time. Lower panel: Power dissipated inside CoRoT-7b vs. the time. The change in the slope of the curve showing the orbital period evolution occurs when $P_{\rm rot} = 2 P_{\rm orb}$ and corresponds to the limit beyond which inertial waves are no longer excited inside the star leading to a remarkable decrease of the tidal dissipation. The same condition corresponds to the resonance excitation of the radial oscillations and the intense dissipation inside the planet. }
              \label{corot7_detail}%
\end{figure}

\begin{figure}
%\hspace*{-7mm}
 \centering{
 \includegraphics[width=8cm,height=10cm,angle=90]{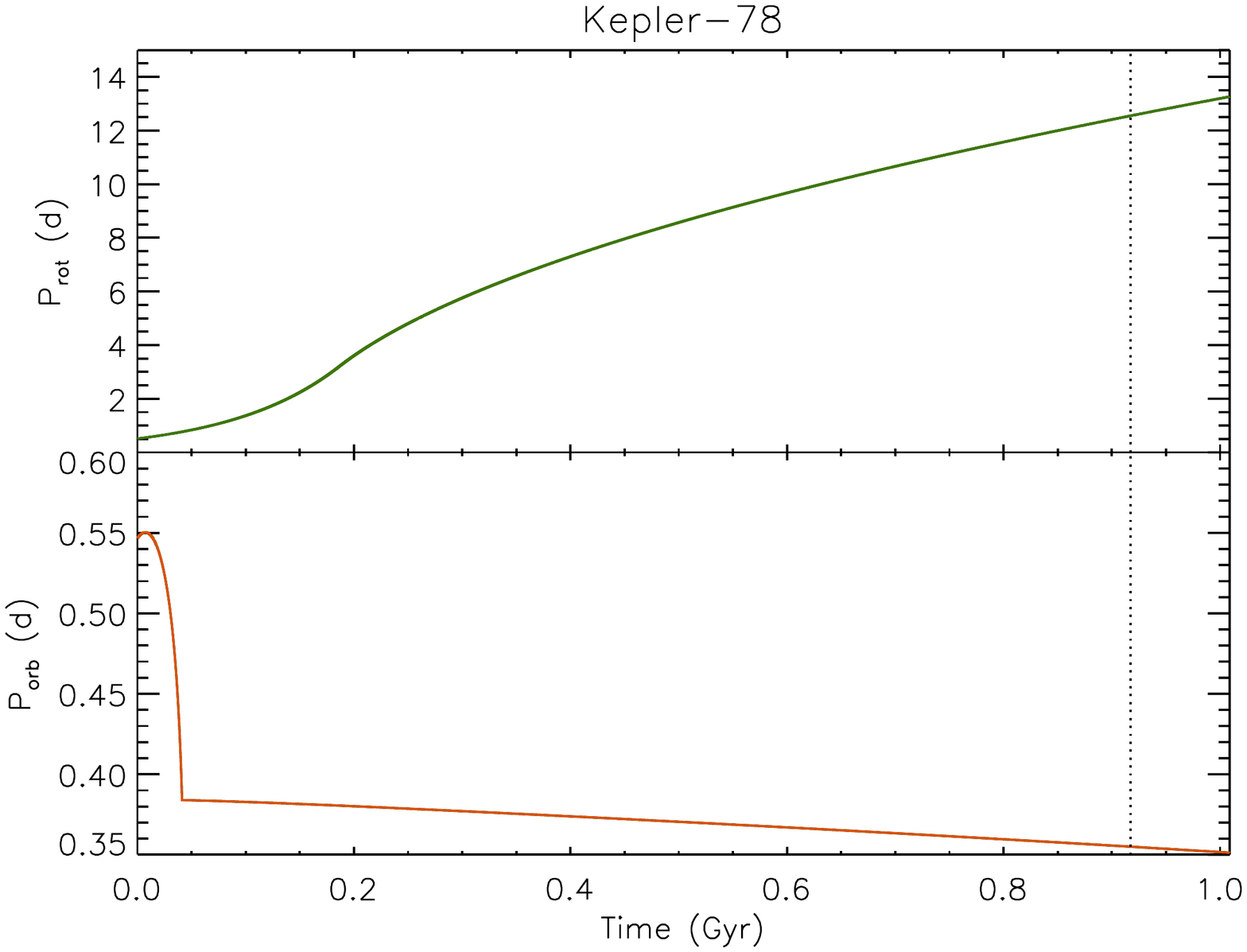}} % Kepler-78_evol_corr.pdf}} % HD106252_1_rhkcorr_KR_final.pdf}}
% \vspace*{-10mm}
   \caption{Same as Fig.~\ref{corot7_evol} for Kepler-78.}
              \label{kepler78_evol}%
\end{figure}

\begin{figure}
%\hspace*{-7mm}
 \centering{
 \includegraphics[width=8cm,height=10cm,angle=90]{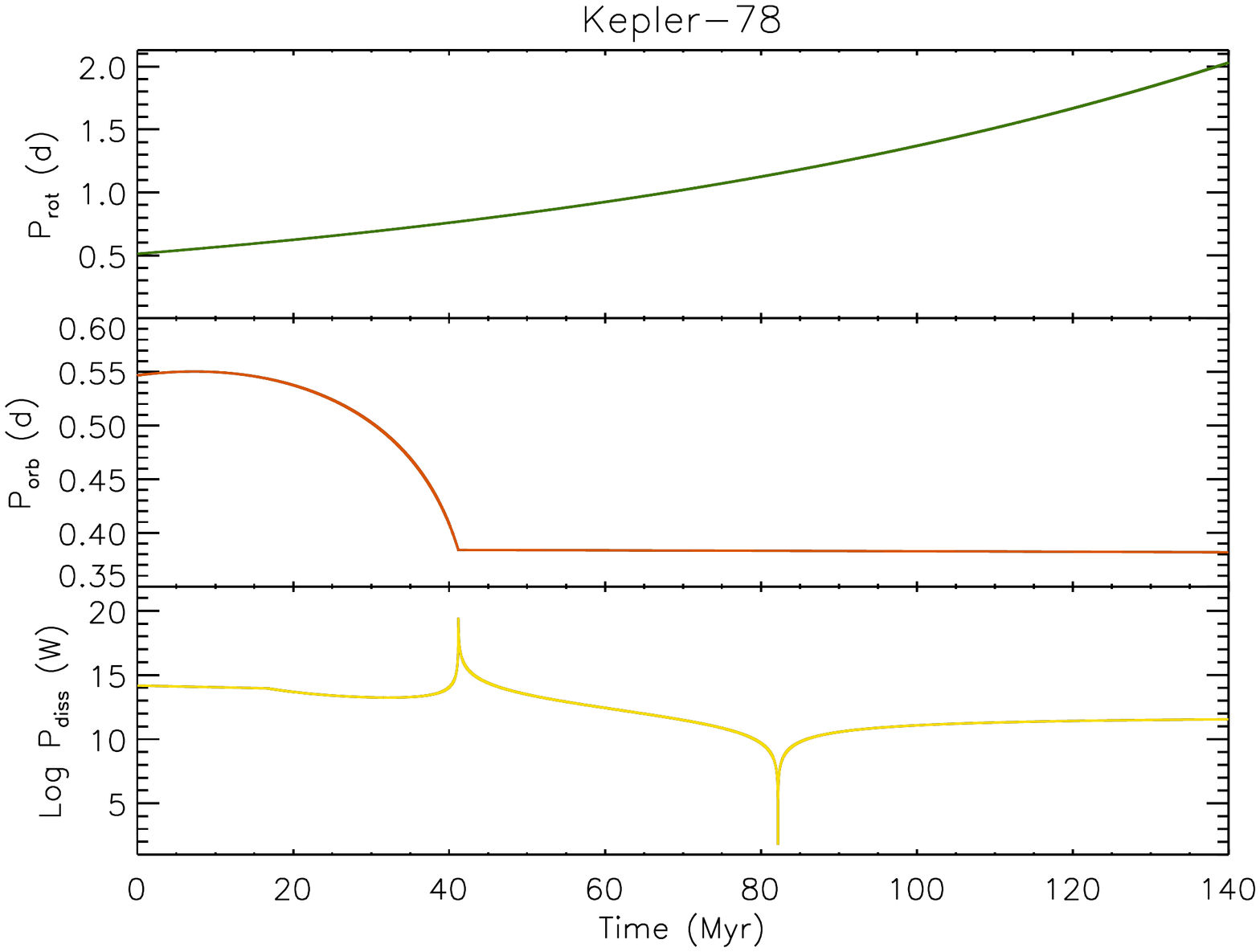}} % Kepler-78_detail_corr.pdf}} % HD106252_1_rhkcorr_KR_final.pdf}}
% \vspace*{-10mm}
   \caption{Same as Fig.~\ref{corot7_detail} for Kepler-78. }
              \label{kepler78_detail}%
\end{figure}

\begin{figure}
%\hspace*{-7mm}
 \centering{
 \includegraphics[width=8cm,height=10cm,angle=90]{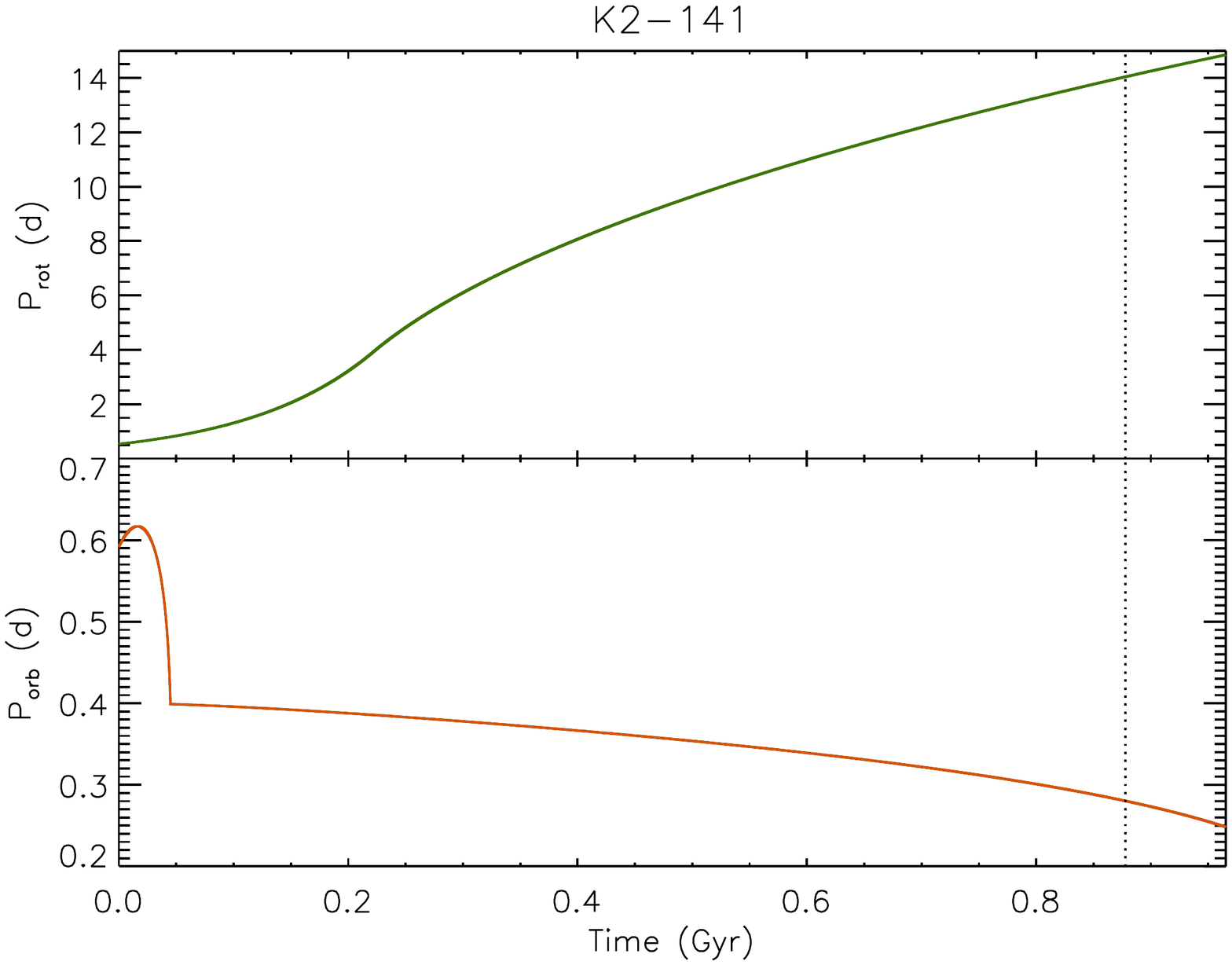}} %K2-141_evol_corr.pdf}} % HD106252_1_rhkcorr_KR_final.pdf}}
% \vspace*{-10mm}
   \caption{Same as Fig.~\ref{corot7_evol} for K2-141.}
              \label{k2-141_evol}%
\end{figure}

\begin{figure}
%\hspace*{-7mm}
 \centering{
 \includegraphics[width=8cm,height=10cm,angle=90]{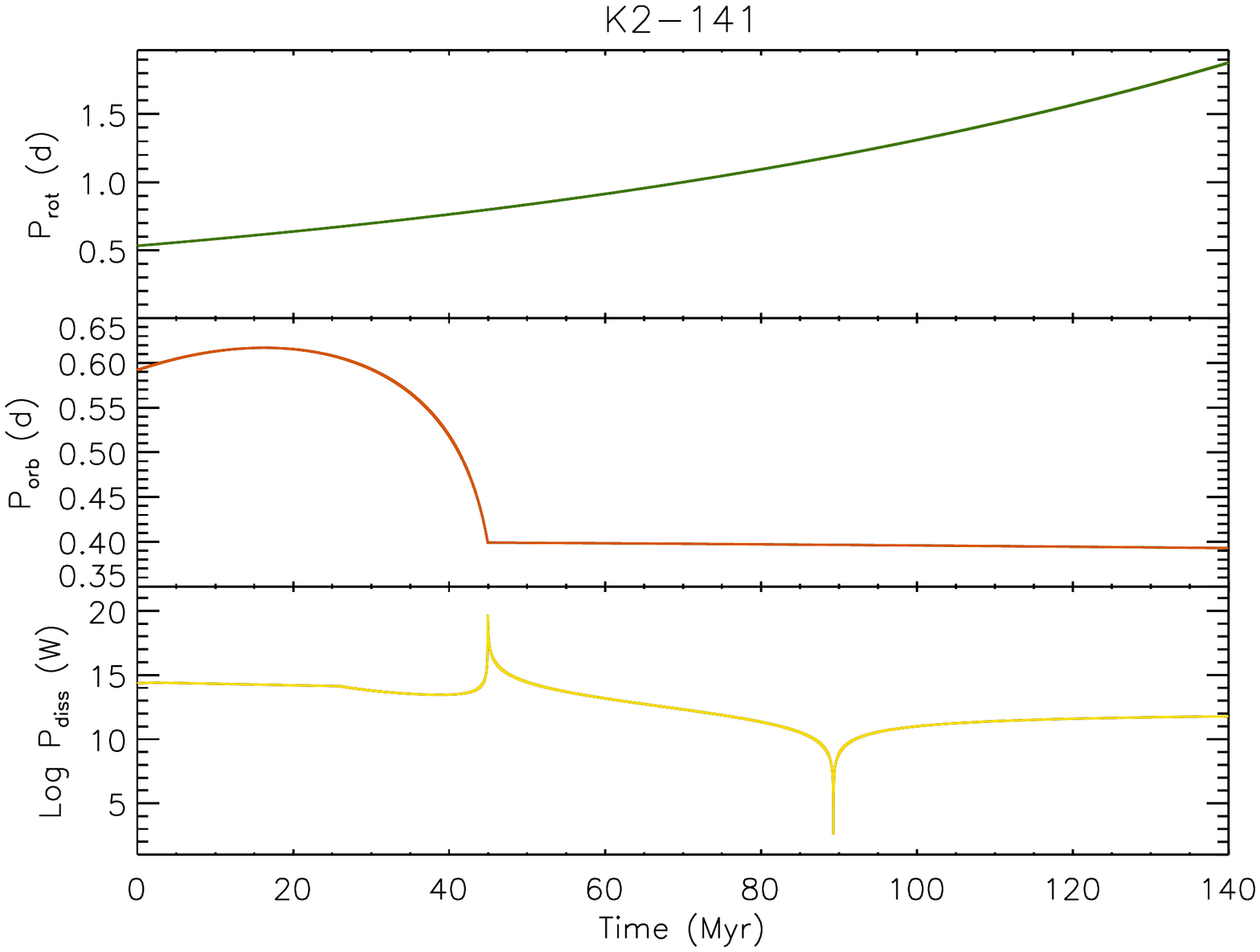}} %K2-141_detail_corr.pdf}} % HD106252_1_rhkcorr_KR_final.pdf}}
% \vspace*{-10mm}
   \caption{Same as Fig.~\ref{corot7_detail} for K2-141. }
              \label{k2-141_detail}%
\end{figure}

For \object{Kepler-78} and \object{K2-141}, the initial rotation period is shorter than the orbital period which leads to an increase of the orbital period and semimajor axis because of the angular momentum transferred by tides from the stellar rotation to the orbital motion during the initial phase of their evolution (cf. Figs.~\ref{kepler78_detail} and~\ref{k2-141_detail}). Conversely, in \object{CoRoT-7}, $P_{\rm rot}$ is initially longer than $P_{\rm orb}$, therefore tides start immediately to decrease the semimajor axis and the orbital period of the planet (cf. Fig.~\ref{corot7_detail}). 

A slope change occurs in the plots of the orbital period vs. the time when the stellar rotation period is twice the orbital period. This corresponds to the transition from the strong tidal coupling  maintained by the excitation of inertial waves inside the star to the regime with only the equilibrium tide. This remarkably reduces the tidal torque inside the star and leads to a much slower decay of the planetary orbit. The excitation of the radial orbital oscillations of the planet occurs close to the transition between these two tidal regimes because the resonance frequency $\gamma_{\rm res} = \sqrt{n^{2} - D^{2}/4}$ is very close to the orbital mean motion $n$, given that $D \ll n$. 

The relative amplitude of the radial oscillations at resonance $\langle x^{2} \rangle^{1/2}$ is also reported in Table~\ref{systems_models}. It is computed from the power spectrum of the forcing as illustrated in Sect.~\ref{sol_rad_motion} (cf. eq.~\ref{x^2m}). Its value is very small, thus verifying our assumption that $| x | \ll 1$ in eq.~(\ref{x_defin}) in Sect.~\ref{radial_motion},  and is extremely difficult to detect because of the short duration of the excitation in comparison to the main-sequence lifetime of the systems (see below). Any residual eccentricity would be damped by the tides inside the planet on a timescale $\tau_{\rm e} = 1/(2b)$ that is shorter than or comparable with a Myr when considering a value of $Q^{\prime \rm T}_{\rm p} = 1425$ as in the case of our Earth \citep{Lainey16}. 

The maximum dissipated power $P_{\rm T \, max}$ is listed in Table~\ref{systems_models} together with the total energy dissipated inside the planet $E_{\rm D}$ and the time needed to cross the resonance, that is, the time interval during which the instantaneously dissipated power is greater than or equal to one half of  the maximum power. The resonance crossing time is very short in comparison with the evolutionary timescales and the timescale needed by sub-solidus convection to transport the heat dissipated inside the mantle of a telluric planet, that is, of the order of $10^{8}$ yr in the case of our Earth. For this reason, we can assume that all the heat dissipated by the proposed mechanism is stored into the planetary interior. The amount of heat available per unit mass of the planet $E_{\rm D}/m_{\rm p}$ is also listed in Table~\ref{systems_models} and it shall be used to infer the effects of the heating in the next Section. 

The heat dissipation produced by our mechanism is restricted to a short time interval of the order of $10^{4}$ yr occurring after $\approx 40-50$~Myr since our host stars have settled on the main-sequence. Another resonance could have occurred during the pre-main-sequence evolution of those systems when the contracting and spinning up protostar could have matched  the condition for the resonance $P_{\rm rot} \simeq 2 P_{\rm orb}$ while evolving towards the ZAMS. For simplicity, we do not consider this possibility and limit ourselves to model only the resonance occurring after the ZAMS. {Finally, we note that when $P_{\rm rot} = 3 P_{\rm orb}$, the amplitude of the forcing term  $A_{\rm s}$ in eq.~(\ref{fforcing_amp1}) becomes zero, thus producing a drop of the dissipated power inside the planet clearly evident in the bottom panels of Fig.~\ref{corot7_detail}, \ref{kepler78_detail}, and~\ref{k2-141_detail}.}
%%%%%%%%%%%%%%%%%%%%%%%%%%%%%%%%%%%%
\begin{table}
\caption{Stellar and planetary parameters adopted for the systems considered in our illustrative applications. They were collected from the Extrasolar Planet Encyclopedia (exoplanet.eu); \citet{Moutouetal16}; \citet{Malavoltaetal18}; and \citet{Barraganetal18}. }
\begin{center}
\begin{tabular}{cccc}
\hline
System & CoRoT-7 & Kepler-78 & K2-141 \\
\hline
$m_{\rm s}$ (M$_{\odot}$) & 0.93 & 0.81 & 0.71 \\ 
$R_{\rm s}$ (R$_{\odot}$)  & 0.87 & 0.74 & 0.68 \\
$m_{\rm p}$ (M$_{\oplus}$) & 4.74 & 1.875 & 5.08 \\
$R_{\rm p}$ (R$_{\oplus}$) & 1.52 & 1.15 & 1.51 \\
$P_{\rm rot}$ (d) & 23.6 & 12.6 & 14.0 \\
$P_{\rm orb}$ (d) & 0.8536  & 0.3550 & 0.2803 \\ 
Age (Myr) & 2400 & 625 & 740 \\
\hline
\end{tabular}
\end{center}
\label{systems_table}
\end{table}
%%%%%%%%%%%%%%%%%%%%%%%%%%%%%%%%%%
%%%%%%%%%%%%%%%%%%%%%%%%%%%%%%%%%%%%
\begin{table*}
\caption{Adopted model parameters and quantities derived for the systems considered in our illustrative applications (see the text). The eccentricity damping timescale $\tau_{\rm e}$ is computed assuming $Q^{\prime \rm T}_{\rm p} = 1425$ in eq.~(\ref{tau_ecc}). }
\begin{center}
\begin{tabular}{cccc}
\hline
System parameter & CoRoT-7 & Kepler-78 & K2-141 \\
\hline
$a( t=0)$ (au) & 0.0177 & 0.0122 & 0.0123 \\
$P_{\rm orb} (t=0) $ (d) & 0.89154 & 0.54666 & 0.59192 \\
$P_{\rm rot} (t=0) = 2\pi/\Omega_{0}$ (d) & 0.93612 & 0.51223 & 0.53273\\
$\Omega_{\rm sat}$ ($\Omega_{\odot}$) & 11.0 & 8.0 & 6.5 \\
$T_{\rm s}/I_{\rm s}$ & $10^{-6}$ & $10^{-6}$ & $10^{-6}$ \\
$(D/2)^{-1}$ (d) & 500 & 500 & 500 \\ 
$Q^{\prime \rm T}_{\rm s}$ (IW) & $3.505 \times 10^{5}$ & $5.247 \times 10^{4}$ & $5.676 \times 10^{4}$ \\
$Q^{\prime \rm T}_{\rm s}$ (ET) & $4.294 \times 10^{8}$ & $7.871 \times 10^{7}$ & $5.960 \times 10^{7}$ \\
Model estimated age (Myr) & 4406 & 917 & 877 \\
$\langle x^{2} \rangle^{1/2}$ & $5.328 \times 10^{-3}$ & $5.820 \times 10^{-3}$ & $4.309 \times 10^{-3}$ \\
$\tau_{\rm e}$ (Myr) & 1.4 & 0.06 & 0.05 \\ 
$P_{\rm T \, max} $ ($10^{19}$~W) & {0.1768} & {2.962} & {5.127} \\
$E_{\rm D}$ ($10^{32}$~J) & {0.075} & {0.6311} & {1.305} \\
$E_{\rm D}/m_{\rm p}$ ($10^{6}$~J/kg) & {0.265} & {5.634} & {4.300}\\ 
Resonance crossing time (yr) & $4.45 \times 10^{4}$ & $1.31 \times 10^{4}$ & $1.17 \times 10^{4}$ \\ 
\hline 
\end{tabular}
\end{center}
\label{systems_models}
\end{table*}
%%%%%%%%%%%%%%%%%%%%%%%%%%%%%%%%%%%%%

%%%%%%%%%%%%%%%%%%%%%%%%%%%%%%%%%%%%%
%%%%%%%%%%%%%%%%%
%%%%%%%%%%%%%%%%%%%%%%%%%%%%%%%%%%%%%
%%%%%%%%%%%%%%%%%
%%%%%%%%%%%%%%%%%%%%%%%%%%%%%%%%%%%%%
%%%%%%%%%%%%%%%%%
%%%%%%%%%%%%%%%%%%%%%%%%%%%%%%%%%%%%%
%%%%%%%%%%%%%%%%%
%%%%%%%%%%%%%%%%%%%%%%%%%%%%%%%%%%%%%
%%%%%%%%%%%%%%%%%
%%%%%%%%%%%%%%%%%%%%%%%%%%%%%%%%%%%%%
%%%%%%%%%%%%%%%%%

\section{Discussion and conclusions}
\label{discussion}
We found that a resonance between the stellar rotation and the orbital motion of an ultra-short-period planet ($P_{\rm orb} \la 1.0$~days) can occur during the first phase of the evolution of the star soon after ($\approx 40-50$~Myr) it reached the ZAMS. The resonance happens when the stellar rotation period is very close to twice the orbital period of the planet. This implies  that the star must reach the ZAMS as a fast rotator with $P_{\rm rot} \la 1-2$~days for the resonance to occur. As a consequence, a  large amount of heat ($\approx 10^{31}-10^{32}$~J) is dissipated in the planetary interior during a short time interval ($(1-5) \times 10^{4}$ yrs), much shorter than the timescale for its transport by conduction or convection in the planetary interior \citep[cf.][]{Jaupartetal07}. 

Immediately after their formation, telluric planets will very likely experience a short phase of complete melting owing to the heat released in their accretion and core formation processes \citep[][]{Jaupartetal07,Chaoetal20}. The duration of this phase is limited to $\approx 10$~Myr for an Earth-mass planet because of the rapid cooling of the interior magma ocean \citep[e.g.][]{Abe97,Jaupartetal07}. Therefore, we can assume that the mantles of our ultra-short-period planets  have had enough time to become nearly solid when they experience the rapid and intense heating produced by the process introduced above, except for their outermost layers where stellar irradiation or magnetic induction can maintain a temperature higher than $\sim 1300-1400$~K sufficient to keep silicates liquid at low pressures \citep{Chaoetal20}. 

The intense and rapid heating produces a huge increase of the temperature in the interior of the planet considering that the specific heat for an Earth-like mantle is  $\sim 1250$~J~kg~K$^{-1}$. From the energy per unit mass listed in Table~\ref{systems_models}, we estimate a temperature increase of $\sim 10^{3}-10^{4}$~K, that is sufficient {in several cases} to completely melt the planetary interior even including a latent heat of mantle melting of $\sim 3.2 \times 10^{5}$~J~kg$^{-1}$ \citep[e.g.][]{Abe97,DriscollBarnes15}, thus resuming the state experienced by the planet immediately after its formation and core differentiation.  The amount of energy $E_{\rm D}$ dissipated into the planet does not depend on its tidal quality factor $Q^{\prime \rm T}_{\rm p}$, therefore this conclusion appears to be robust. 

{The peak power reached by the dissipation mechanism we introduced is of the order of $10^{19}$~W} and is huge in comparison with the present internal heat flow in our Earth that is $\sim 47 \pm 2$~TW \citep{DaviesDavies10}. The internal heat flux coming from the decay of radioactive nuclids is presently estimated to be $\sim 22$~TW \citep{ONeilletal20}, although the recent detection of geoneutrinos would suggest a larger value of $\sim 38 \pm 14$~TW \citep{Agostinietal20}. An extrapolation to an age of about $\sim 30-40$~Myr, indicates a contribution of the radioactive heating between $100$ and $200$~TW, thus much smaller than the heat power released by our mechanism. The internal heat released during the process of the planet formation and the core differentiation is assumed to be dissipated for the most part during the first $10-30$~Myr of the planet evolution, therefore, we may assume that the mean temperature of the mantle of an average telluric planet at an age of $40-50$~Myr is only about $300-400$~K higher that the present Earth value \citep{Jaupartetal07} implying a cooling power of the order of $30-50$~TW at those young ages. Note that the present power due to the cooling of the interior of our Earth is $\approx 25$~TW, including $\approx 8$~TW coming from the slow solidification of the inner metallic core of our planet \citep[e.g., Table~11 of][]{Jaupartetal07}. 

The magnetic interaction between a telluric planet endowed with an intrinsic magnetic field and its young host star can lead to a  release of power up to $10^{19}$~W through magnetic reconnection \citep[e.g.,][]{Lanza12} or up to $10^{21}$~W in the case of the energy stored by Maxwell stresses in an interconnecting magnetic loop \citep[e.g.,][]{Lanza13}. However, such powers are released outside the planetary body at the boundary of its magnetosphere or close to its surface, respectively, because the intrinsic planetary field and the high electric conductivity of the ionized planetary atmosphere shield the planetary interior preventing the penetration of the stellar field inside the planet. Therefore, such kinds of dipolar magnetic interactions cannot heat the planetary interior and affect its evolution. The situation is different in the so-called unipolar inductor regime, when the planet has no significant intrinsic magnetic field and has a low, although finite, electric conductivity in its mantle, because in that case the stellar field can penetrate into its interior. In such a regime, the time variation in the stellar magnetic flux across the section of the planet can dissipate powers up to $\approx 10^{22}-10^{23}$~W in the case of a young rapidly rotating star \citep[e.g.,][]{LaineLin12}, which  can exceed the peak power dissipated by the mechanism proposed in the present paper.  %}

The dissipation predicted by our mechanism mainly occurs in the planetary layers most affected by the tidal deformation, which implies that most of the heating is concentrated in the outer half of the planet. Therefore, it can profoundly affect the cooling of the planetary core, for instance, inhibiting its convective motions and shutting off the operation of the planet hydromagnetic dynamo until the previous temperature gradient is re-establish which may imply timescales of the order of at least $10^{7}-10^{8}$~yr. This may have a remarkable effect on the planetary atmosphere that can be completely eroded by the stellar high-energy radiation and wind, given that the protection by the planetary magnetic field has ceased. {Moreover, the shutting off of the planetary dynamo may change the regime of the star-planet magnetic interaction to the unipolar inductor. Such a regime, characterized by a huge internal dissipation that persists for several tens of Myr until the stellar field weakens owing to the braking of stellar rotation, can deeply affect the evolution of the planetary interior and of its atmosphere \citep[cf.][]{LaineLin12}. }

From a more general point of view, a phase of very intense heating as predicted by our model (or the unipolar inductor model) can lead to changes in the chemistry of the interior with consequences for the properties of the surface and/or of the secondary planetary atmosphere that could be detectable, at least in principle. We do not explore these consequences here because they are outside the scope of this work that it devoted to introduce the physical basis of this new heating mechanism. The evolution of a global magma ocean and its interaction with a planetary atmosphere leading to alterations of a  planet's observable properties are discussed in, e.g., \citet{Chaoetal20}. 

Given that not all late-type dwarf stars begin their ZAMS evolution as fast rotators and that ultra-short-period planets may not be already in place during the early phases of their main-sequence evolution, not all these planets may experience this phase of early and highly intense heating. A future analysis of the surface properties and of the secondary atmospheres of ultra-short-period planets may reveal the signatures of that phase, thus confirming the prediction about such a new internal heating mechanism. In any case, the main result of the present work suggests that some ultra-short-period planets may have had an evolution remarkably different from that of their siblings on longer period orbits or that migrated close to their host stars after the first $\approx 100$~Myr of their main-sequence lifetime.  

\begin{acknowledgements}
The author gratefully acknowledges an anonymous referee for a careful reading of the manuscript and several valuable comments that helped to improve this work.  This investigation has been supported by the PRIN-INAF 2019 "Planetary systems at young ages (PLATEA)",  the PI of which is Dr.~S.~Desidera. The use of the MESA-Web interface in December 2020 to compute illustrative stellar models is also acknowledged. 
\end{acknowledgements}

\appendix
\section{Computation of some integrals with the method of the residues}
\label{math_app}
To compute the integrals appearing in eqs.~(\ref{pt2}) and~(\ref{ed}) we make use of the theorem of the residues  \citep{Smirnov64}. Let us consider a complex rational function $f(z)$ of a complex argument $z \in \mathbb{C}$ defined as 
\begin{equation}
f(z) = \frac{\phi(z)}{\psi(z)},
\end{equation}
where $\phi(z)$ and $\psi(z)$ are complex polynomials and the degree of $\psi(z)$ is greater than the degree of $\phi(z)$ by at least two, while $\psi(z)$ has no zeros on the real axis. 

\begin{figure}
%\hspace*{-7mm}
 \centering{
 \includegraphics[width=9cm,height=6cm,angle=0,trim=105 105 105 105,clip]{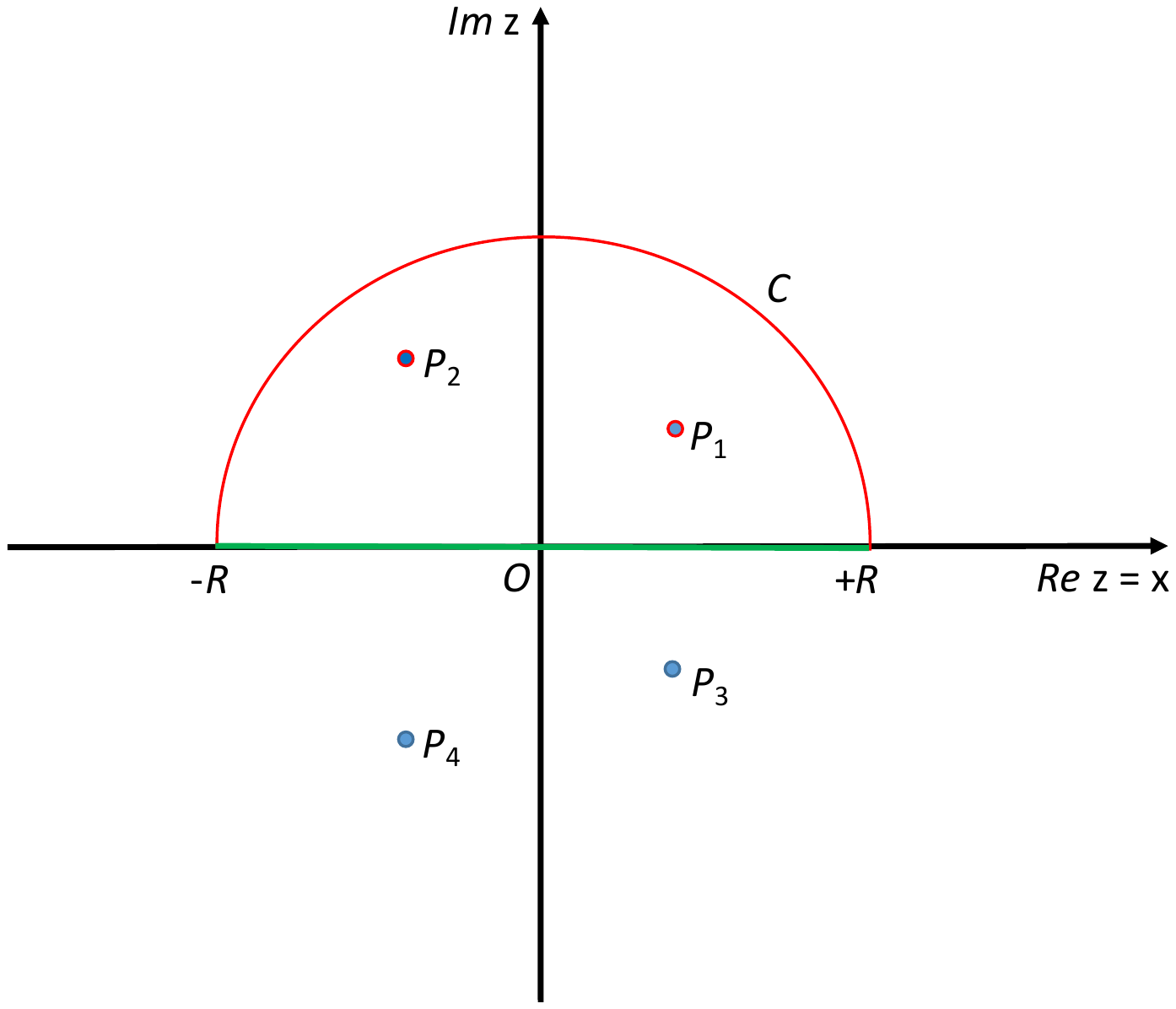}} %figure3.pdf}} % HD106252_1_rhkcorr_KR_final.pdf}} 60 60 62 62 
% \vspace*{-10mm}
   \caption{The complex plane with the semicircle $C$ (in red) and its diameter, that is, the segment of the real axis $[-R, +R]$ (in green). Their union forms the closed contour $\Sigma$ along which the complex function $f(z)$ is integrated in eq.~(\ref{theor_res}). For the sake of illustration, we show two poles of the complex integrand function $P_{1}$ and $P_{2}$ (in blue encircled in red), inside the contour $\Sigma$ and two poles outside the contour $\Sigma$. Only the residues in $P_{1}$ and $P_{2}$ are to be  considered in the computation of the integral of $f(z)$ along $\Sigma$ according to the theorem of the residues, while the residues in the poles $P_{3}$ and $P_{4}$ (in blue), lying outside $\Sigma$, are not to be included in the summation in eq.~(\ref{theor_res}).   }
              \label{semicircle}%
\end{figure}

Let us consider the integral of $f(z)$ along the closed curve $\Sigma$  consisting of the union of the semicircle $C$ in the positive imaginary semiplane having its centre at the origin $O$ and radius $R$ with its diameter, that is, the segment of the real axis $[-R, R]$ (see Fig.~\ref{semicircle}). The theorem of the residues states that \citep[][\S~58]{Smirnov64}
\begin{equation}
\int_{\Sigma} f(z) \, dz = 2\pi i \sum_{k} r_{k}, 
\label{theor_res}
\end{equation}
where $i = \sqrt{-1}$ and the $r_{k}$ are the residues of $f(z)$  in the poles enclosed by the closed curve $\Sigma$. We can write the integral along $\Sigma$ as the sum of the integral along the segment on the real axis plus the integral along $C$ as
\begin{equation}
\int_{\Sigma} f(z) \, dz = \int_{-R}^{+R} f(x) \, dx + \int_{C} f(z) \, dz,
\end{equation}
where $f(x)$ is the restriction of the complex function $f(z)$ to the real axis. 
If we make $R \rightarrow \infty$, the integral along $C$ tends to zero because the function $f(z)$ tends to zero at least as $z^{-2}$ for $z \rightarrow \infty$ and we can apply eq.~(\ref{theor_res}) to find:
\begin{equation}
\int_{-\infty}^{\infty} f(x) \, dx = 2\pi i \sum_{\Im{z_{k}} > 0} r_{z_{k}},
\label{theor_res1}
\end{equation}
where the summation is extended to all the residues of $f(z)$ in the upper complex semiplane, that is, to all and only the residues $r_{z_{k}}$ in the poles $z_{k}$ of $f(z)$ whose  imaginary parts  are  positive. The improper integral in eq.~(\ref{theor_res1}) exists because $f(x)$ goes to zero as $x^{-2}$ when $x\rightarrow \pm \infty$ and its denominator has no zeros on the real axis. 

Since $f(z)$ is the ratio of two polynomials, its poles are the zeros of the denominator $\psi(z)$. In all the cases considered in this Appendix, these are simple poles (that is, roots of the first order of the polynomial $\psi(z)$), therefore, the residual in a pole $z_{k}$ is given by 
\begin{equation}
r_{k} = \left[ f(z) ( z- z_{k}) \right]_{z=z_{k}} = \frac{\phi(z_{k})}{\psi^{\prime} (z_{k})},
\label{res_exp}
\end{equation}
where $\psi^{\prime}(z)$ is the derivative of the polynomial $\psi(z)$ \citep[][\S~21]{Smirnov64}. 
%%%%%%%%%%%%%%%%%%%%%%%%%%%%%%%%%%%%%
%%%%%%%%%%%%%%%%%

\subsection{Computation of the integral in eq.~(\ref{pt2})}
In this case, $\phi(z) = z^{2}$ and $\psi(z) = (n^{2}-z^{2})^{2} + 4b^{2} z^{2}$, where $z$ is a complex argument, while $n$ and $b$ are real numbers.  By differencing the latter, we find $\psi^{\prime} (z) = 4 (z^{2} + 2 b^{2} - n^{2}) z$. The poles are given by $\psi(z)=0$, that is:
\begin{eqnarray}
\nu_{1,2} = ib \pm \sqrt{n^{2} - b^{2}}, \\
\nu_{3, 4} = \nu_{1,2}^{*} = -ib \pm \sqrt{n^{2} - b^{2}}.
\end{eqnarray}
From these equations, it immediately follows that $\nu_{2} = -\nu_{1}^{*}$ and $\nu_{4} = - \nu_{3}^{*}$, where the asterisk indicates complex conjugation. 

The two poles in the positive imaginary semiplane are $\nu_{1, 2}$. We have $\phi(\nu_{2}) = [\phi(\nu_{1})]^{*}$, while $\psi^{\prime} (\nu_{2}) = - [\psi^{\prime} (\nu_{1})]^{*}$. Considering eq.~(\ref{res_exp}), this implies that the residues in $\nu_{1}$ and $\nu_{2}$ are related by $r_{\nu_{2}} = - r_{\nu_{1}}^{*}$. Using this result and after some algebra, we find that the sum of the residues in eq.~(\ref{theor_res1}) is:
\begin{equation} 
\sum_{\Im{\nu_{k}} > 0} r_{\nu_{k}} = r_{\nu_{1}} + r_{\nu_{2}} = 2 i \, \Im{r_{\nu_{1}}} = -\frac{i}{4b}, 
\end{equation}
where $\Im z$ denotes the imaginary part of the complex number $z$. 
By applying eq.~(\ref{theor_res1}), finally we find
\begin{equation}
\int_{-\infty}^{\infty}  \frac{ \nu^{2} d\nu}{(n^{2}-\nu^{2})^{2} + 4 b^{2} \nu^{2}} = \frac{\pi}{2b}. 
\end{equation}
A similar integral, that is useful to compute $\langle x^{2} \rangle$ for our applications (cf. eq.~\ref{x^2m}), can be evaluated with the same method:
\begin{equation}
\int_{-\infty}^{\infty}  \frac{d\nu}{(n^{2}-\nu^{2})^{2} + 4 b^{2} \nu^{2}} = \frac{\pi}{2bn^{2}}.
\end{equation}

\subsection{Computation of the integral in eq.~(\ref{ed})}

The integral in eq.~(\ref{ed}) can be cast in the form
\begin{equation}
\int_{-\infty}^{\infty} \frac{a+ \gamma^{2}}{\left(  b +\gamma^{2} \right)^{2} + c} \, d\gamma, 
\end{equation}
where  $a = n^{2} + D^{2}/4 > 0$, $b = -n^{2} + D^{2}/4$, and $c = a^{2} - b^{2} = D^{2} n^{2} >0$, are real numbers. 

In this case, the functions $\phi(z) = a +z^{2}$ and $\psi(z) =  (b + z^{2})^{2} + c$. The poles of the complex integrand are the four roots  of $\psi(z) = 0$. To find their expressions, we consider the complex number $w = -b + i \sqrt{c}$. Its trigonometric expression is $w = \rho \exp(i\theta)$, where $\rho = \sqrt{b^{2} + c} = a >0$, $\cos \theta = -b/a$, and $\sin \theta = \sqrt{c}/a$. 

The four roots of $\psi(z) = 0$ are $\gamma_{1, 2} = \sqrt{w}$ and $\gamma_{3,4} =  \sqrt{w^{*}}$. The square root of $w$ can be obtained from its trigonometric expression as $\sqrt{w} = \sqrt{a} \exp [i(\theta + 2 k\pi)/2]$, where $k$ is an integer. Taking into account that $\sin (\theta/2) = \sqrt{(1-\cos \theta)/2}$ and $\cos (\theta/2) = \sqrt{(1+ \cos \theta)/2}$, we find the four poles of $f(z)$ 
\begin{eqnarray}
\gamma_{1,2} = \pm \frac{1}{\sqrt{2}} \left( \sqrt{a-b} + i \sqrt{a+b} \right), \nonumber \\
\gamma_{3,4} = \pm \frac{1}{\sqrt{2}} \left( \sqrt{a-b} - i \sqrt{a+b} \right). 
\end{eqnarray}
The two poles in the positive imaginary semiplane are $\gamma_{1}$ and $\gamma_{4}$. They are related by $\gamma_{1} = - \gamma_{4}^{*}$ and, by applying eq.~(\ref{res_exp}), where $\psi^{\prime}(z) = 4z (b +z^{2})$, this implies  $r_{\gamma_{1}} =- r_{\gamma_{4}}^{*}$. Therefore, the sum of the residues in the positive imaginary semiplane is:
\begin{equation}
\sum_{\Im{r_{\gamma_{k}}}> 0} r_{\gamma_{k}} = r_{\gamma_{1}} + r_{\gamma_{4}} = 2i \Im{r_{\gamma_{1}}} = - \frac{\sqrt{2}}{2}i {\frac{1}{\sqrt{a+b}}}. 
\end{equation}
Finally, by applying the theorem of the residues (eq.~\ref{theor_res1}), we find
\begin{equation}
\int_{-\infty}^{\infty} \frac{a+\gamma^{2}}{\left(b+\gamma^{2} \right)^{2} + c} \, d\gamma = \frac{\pi \sqrt{2}}{\sqrt{a+b}}.
\end{equation}
In conclusion, we find that
\begin{equation}
\int_{-\infty}^{\infty} \frac{A_{\rm s}^{2} D}{2\sqrt{2 \pi}} \frac{\frac{1}{4} D^{2} + \gamma^{2} + \nu^{2}}{\left( \frac{1}{4} D^{2} + \gamma^{2} - \nu^{2} \right)^{2} + D^{2} \nu^{2}} \, d\gamma = \frac{\sqrt{2\pi}}{2} A_{\rm s}^{2}. 
\label{energy_integral}
\end{equation}

%\appendix
%\section{Computation of some useful integrals}
%\label{math_app}

\end{document}